\documentclass[a4paper]{article}

\usepackage{amssymb,amsmath,amsthm}
\usepackage[totalwidth=17cm,totalheight=24cm]{geometry}
\usepackage{graphicx,verbatim}

\newcommand{\C}{{\mathbb C}}
\newcommand{\R}{{\mathbb R}}
\newcommand{\Z}{{\mathbb Z}}
\newcommand{\Hq}{{\mathbb H}}
\newcommand{\Oc}{{\mathbb O}}

\newcommand{\id}{{\mathbb I}}
\newcommand{\im}{{\rm i\,}}

\newcommand{\x}{{\bf x}}

 \theoremstyle{plain}
  \newtheorem{theorem}{Theorem}
   \newtheorem{proposition}[theorem]{Proposition}
  \newtheorem{lemma}[theorem]{Lemma}

  \newtheorem{manualtheoreminner}{Theorem}
\newenvironment{manualtheorem}[1]{%
  \renewcommand\themanualtheoreminner{#1}%
  \manualtheoreminner
}{\endmanualtheoreminner}

  \theoremstyle{definition}
  \newtheorem{definition}[theorem]{Definition}

  \theoremstyle{remark}
  \newtheorem{remark}{Remark}

\newcommand{\be}{\begin{eqnarray}}
\newcommand{\ee}{\end{eqnarray}}

\begin{document}
 \pagestyle{plain}
\title{Geometry of ${\rm Spin}(10)$ Symmetry Breaking}

\author{Kirill Krasnov \\ {}\\
{\it School of Mathematical Sciences, University of Nottingham, NG7 2RD, UK}}

\date{v2: January 2023}
\maketitle

\begin{abstract}\noindent We provide a new characterisation of the Standard Model gauge group $G_{\rm SM}$ as a subgroup of ${\rm Spin}(10)$. The new description of $G_{\rm SM}$ relies on the geometry of pure spinors. We show that $G_{\rm SM}\subset{\rm Spin}(10)$ is the group that stabilises a pure spinor $\Psi_1$ and projectively stabilises another pure spinor $\Psi_2$, with $\Psi_{1,2}$ orthogonal and such that their arbitrary linear combination is still a pure spinor. Our characterisation of $G_{\rm SM}$ relies on the facts that projective pure spinors describe complex structures on $\R^{10}$, and the product of two commuting complex structures is a what is known as a product structure. For the pure spinors $\Psi_{1,2}$ satisfying the stated conditions the complex structures determined by $\Psi_{1,2}$ commute and the arising product structure is $\R^{10}=\R^{6}\oplus\R^4$, giving rise to a copy of Pati-Salam gauge group inside ${\rm Spin}(10)$. Our main statement then follows from the fact that $G_{\rm SM}$ is the intersection of the Georgi-Glashow ${\rm SU}(5)$ that stabilises $\Psi_1$, and the Pati-Salam ${\rm Spin}(6)\times{\rm Spin}(4)$ arising from the product structure determined by $\Psi_{1,2}$. We have tried to make the paper self-contained and provided a detailed description of the creation/annihilation operator construction of the Clifford algebras ${\rm Cl}(2n)$ and the geometry of pure spinors in dimensions up to and including ten. 
\end{abstract}

\section{Introduction}

${\rm Spin}(10)$ is a very compelling grand unification scheme, encompassing both the Georgi-Glashow ${\rm SU}(5)$ and the Pati-Salam ${\rm SU}(2)\times{\rm SU}(2)\times{\rm SU}(4)$ GUTs. Similar to the ${\rm SU}(5)$ unification, ${\rm Spin}(10)$ is simple. One of its main attractive features is the fact that it unifies all fermionic states of one generation in a single irreducible (Weyl spinor) representation of ${\rm Spin}(10)$.

The weakness of the ${\rm Spin}(10)$ GUT lies in the fact that it is far from obvious how it breaks down to the Standard Model gauge group $G_{\rm SM}={\rm SU}(3)\times{\rm SU}(2)\times{\rm U}(1)_Y$, or to the unbroken in Nature gauge group ${\rm SU}(3)\times{\rm U}(1)_Q$. There are many possible symmetry breaking patters, and the arising complexity is well-illustrated by e.g. Figure 1.2 in \cite{DiLuzio:2011mda}. The simplest and most studied symmetry breaking scenario is to consider two Higgs fields, one in the spinor representation ${\bf 16}$, and one in the adjoint ${\bf 45}$ of ${\rm Spin}(10)$. Choosing the spinor Higgs to be a multiple of what mathematicians know as the pure spinor breaks the symmetry to ${\rm SU}(5)$. Then further choosing the adjoint Higgs to be appropriately aligned with the spinor Higgs breaks the symmetry to $G_{\rm SM}$. To break the symmetry further down to ${\rm SU}(3)\times{\rm U}(1)_Q$ one needs another Higgs field, usually taken to be in the vector representation $\bf 10$. However, as Figure 1.2. of \cite{DiLuzio:2011mda} illustrates, this is really only one of the many possible scenarios. And the difficulty of the ${\rm Spin}(10)$ GUT is that after a half a century of work on it none of the scenarios studied emerged as the compelling one, see e.g. \cite{DiLuzio:2011mda} for the discussion of some of the arising issues. 

In this situation it maybe justified to step back and approach the question of symmetry breaking from a different perspective. This is what the present article attempts. To motivate our considerations we start by reminding the reader the notion of a geometric structure. We do not attempt to be precise in this discussion, our purpose here is only to motivate what follows. Let $M$ be a smooth manifold of dimension $n$. Then the geometric structure is the reduction of the structure group of the frame bundle of $M$ from ${\rm GL}(n,\R)$ to some of its subgroups. Thus, a metric is the reduction to the orthogonal group ${\rm O}(n)$, a complex structure (when $n=2m$) is the reduction to ${\rm GL}(m,\C)$, etc. It is clear that we can rephrase the notion of the geometric structure in terms of "symmetry breaking", and it is clear that choosing a geometric structure (on a smooth manifold) "breaks the symmetry". The more geometric structures one chooses, the further is the symmetry breaking. 

It may then be worth to reformulate the question of the symmetry breaking from ${\rm Spin}(10)$ to $G_{\rm SM}$ as the question of what are the geometric structures that are chosen to effect such a reduction of symmetry. This is precisely the angle we take in the present paper. This leads us to a new characterisation of the Standard Model gauge group inside ${\rm Spin}(10)$. Thus, we first show that the former arises if one chooses a complex structure on $\R^{10}$, a holomorphic top form (with both data encoded in a choice of a pure spinor), and a suitably aligned with this data product structure $\R^{10}=\R^6\times\R^4$. By itself this characterisation is not new, and it is noticed for example in \cite{Baez:2009dj} that the SM gauge group arises as the intersection of the Georgi-Glashow ${\rm SU}(5)$ and the Pati-Salam ${\rm SU}(2)\times{\rm SU}(2)\times{\rm SU}(4)$, when the two are properly aligned. The novelty of this paper comes from the observation that we can parametrise the required Pati-Salam subgroup by {\bf another} spinor Higgs field. This is because, as we discuss in details in Section \ref{sec:prod-struct}, a product structure that is compatible with a complex structure $J_1$ is nothing else but the product $J_1 J_2$ of two commuting complex structures. A similar construction is at play in the context of bi-Hermitian geometry, see e.g. \cite{Gualtieri:2003dx}, Section 6. Here we propose to use the geometry of two commuting complex structures to parametrise the ${\rm Spin}(10)$ symmetry breaking. Thus, the Pati-Salam subgroup of ${\rm Spin}(10)$ that is aligned with a chosen ${\rm SU}(5)$ (that stabilises a complex structure $J_1$ on $\R^{10}$) can be parametrised by another complex structure $J_2$, provided $J_1, J_2$ commute and the arising product structure is of the type $\R^{10}=\R^6\times\R^4$. Given that the second complex structure on $\R^{10}$ can be parametrised by a projective pure spinor implies that we can encode the data about both ${\rm SU}(5)$ and ${\rm SU}(2)\times{\rm SU}(2)\times{\rm SU}(4)$ into the choice of two pure spinors $\Psi_{1,2}$. The only question is then what are the conditions necessary to guarantee that the arising product structure is of the Pati-Salam type. It is not difficult to answer this question with the right tools, and the answer is provided by our main Theorem \ref{thm:A}, see Section \ref{sec:pair-pure-spin}.

The organisation of this paper is as follows. We start by reviewing, in Section \ref{sec:2n}, the creation/annihilation operator construction of the Clifford algebras ${\rm Cl}(2n)$. We remind here how spinors can viewed as inhomogeneous differential forms (that we call polyforms) in $\C^n$. We also discuss here how the possible reality conditions operators are represented in the language of creation/annihilation operators, and discuss the relation between pure spinors and complex structures. We end this section with the example of ${\rm Cl}(4)$ that illustrates all the constructions. We continue, in Section \ref{sec:oct}, with the description of the models of Clifford algebras ${\rm Cl}(-6), {\rm Cl}(8)$ and ${\rm Cl}(10)$ based on octonions. We describe general facts about pure spinors and associated geometry in Section \ref{sec:pure}. Here we also discuss the classification of spinors in dimensions up to and including ten. We review the basics of GUT models in Section \ref{sec:GUT}. In particular, this section describes how the Standard Model gauge group is the intersection of the Georgi-Glashow and Pati-Salam subgroups of ${\rm Spin}(10)$. Section \ref{sec:particles} describes the explicit dictionary between the elementary particles and polyforms, and elementary particles and octonions. Finally, in Section \ref{sec:new} we describe the main new statement of the paper and prove the main theorem. We conclude with a discussion, which contains a concrete field theory model that is based on the ideas described in this paper.

\section{Creation/annihilation operator model of ${\rm Cl}(2n)$}
\label{sec:2n}

The purpose of this section is review the construction of the Clifford algebras ${\rm Cl}(2n)$ that utilises fermionic creation-annihilation operators. The fact that a model of this sort is possible, and gives a very efficient concrete description of ${\rm Cl}(2n)$, is related to the representation theoretic fact that the reduction of the spin representation $S$ of ${\rm Spin}(2n)$ to the subgroup ${\rm U}(n)\subset{\rm Spin}(2n)$ coincides with the space of inhomogeneous differential forms on $\C^n$, i.e. 
\be\label{spinors-polyforms}
S\Big|_{{\rm U}(n)} = \Lambda \C^n.
\ee

The logic of our exposition in this section is as follows: We first remark that the Clifford generators can be realised as appropriate linear combinations of $n$ pairs of (fermionic, i.e. anti-commuting) creation/annihilation operators. We then provide a realisation of these creation/annihilation operators as those acting in the space of inhomogeneous differential forms in $\Lambda\C^n$. Differential forms thus form a module for the Clifford algebra, and are Dirac spinors. We then explain how the creation/annihilation operator construction depends on a choice of the complex structure on $\R^{2n}$ and in turn determines such a choice. The exposition follows with a discussion of possible "reality" conditions on spinors, i.e. possible anti-linear operators that either commute or anti-commute with all the Clifford generators. We give a quick discussion of pure spinors, and end this Section with an example of ${\rm Cl}(4)$ treated using our creation/annihilation operator approach.

The material in this section is standard. Possible references are \cite{Budinich:1987jzj}, \cite{Budinich:1989bg} and \cite{Harvey} for a more mathematical exposition. A recent reference is \cite{Bhoja:2022dsm} from where we take some of the material.

\subsection{Creation/annihilation operator model}

Given a vector space $V$ equipped with a metric $g$, the Clifford algebra ${\rm Cl}(V)$ is the algebra generated by $V$ subject to the relation $XY + YX = 2g(X,Y) \id$, for arbitrary $X,Y\in V$. 

We now take $V=\R^{2n}$ equipped with the standard definite metric. We write ${\rm Cl}(2n)$ for the corresponding Clifford algebra. It can be realised by introducing $n$ pairs of (fermionic) creation/annihilation operators that satisfy the following anti-commutator relations
\be\label{anti-comm}
\{ a_i, a_j^\dagger \} = \delta_{ij}, \qquad i=1,\ldots,n.
\ee
We then construct the Clifford generators as the appropriate linear combinations of $a_i, a_i^\dagger$
\begin{equation}\label{Gamma-a}
\Gamma_{i+n} := 
a_i + a_i^\dagger ,\qquad 
    \Gamma_{i} := i (a_i - a_i^\dagger), \qquad 
    1 \leq i \leq n.
\end{equation}
It is then easy to verify that the Gamma matrices satisfy the Clifford algebra relations
\begin{equation}\label{clifford relation 2n}
    \Gamma_A\Gamma_B+\Gamma_B\Gamma_A=2\delta_{AB}, \qquad 1 \leq A,B \leq 2n.
\end{equation}
\begin{remark}
Our conventions for Clifford algebras are opposite to those used in most of the mathematical literature, in that the Clifford generators of ${\rm Cl}(2n)$ square to plus the identity rather than minus the identity. So, for us the Clifford algebra where generators square to minus the identity would be denoted by ${\rm Cl}(-2n)$. This other Clifford algebra is also easily described by the creation/annihilation operator construction. Indeed, now the $\Gamma$-matrices given by the sum of the creation and annihilation operators would contain the factor of the imaginary unit, and the ones given by the difference would not. The difference between ${\rm Cl}(2n)$ and ${\rm Cl}(-2n)$ is important in the world of Clifford algebras, but becomes immaterial once the corresponding Spin group is considered. 
\end{remark}
\begin{remark} As the vector space, the Clifford algebra ${\rm Cl}(V)$ is isomorphic to $\Lambda V$, the space of anti-symmetric tensors in tensor powers of $V$. The identification with $\Lambda V$ introduces the natural inner product in ${\rm Cl}(V)$. Thus, if $e_1,\ldots, e_{2n}$ is a natural basis in $V$, then $e_{i_1} \wedge \ldots \wedge e_{i_p}$ is an orthonormal basis for ${\rm Cl}(V)$. The square norm determined by this inner product on ${\rm Cl}(V)$ will be denoted by $||\cdot||$. We will only use the norm on the Clifford algebra to define the notion of the Clifford group in the next subsection. It will not be used in anything that follows. 
\end{remark}

\subsection{Groups ${\rm Pin}(2n)$ and ${\rm Spin}(2n)$}

We remind the standard definitions, see e.g. \cite{Harvey}, Chapter 10. The group ${\rm Cl}^*(V)$ is the group of invertible elements of ${\rm Cl}(V)$. The inverse is defined as $u^{-1} := u/||u||$. 
\begin{definition}
The Pin group is the subgroup of ${\rm Cl}^*(V)$ generated by unit vectors in $V$:
\begin{equation}
{\rm Pin} = \{ a\in {\rm Cl}^*(V): a= u_1 \ldots u_r, \quad u_i\in V \quad \text{and} \quad ||u_i||=1\}.
\end{equation}
\end{definition}
\begin{definition}
The Spin group is the subgroup of ${\rm Cl}^*(V)$ generated by an even number of unit vectors in $V$:
\begin{equation}
{\rm Spin} = \{ a\in {\rm Cl}^*(V): a= u_1 \ldots u_{2r}, \quad u_i\in V \quad \text{and} \quad ||u_i||=1\}.
\end{equation}
\end{definition}
Both groups act naturally on $V$ by reflections, see \cite{Harvey} for more details. The Pin group double covers the orthogonal group ${\rm O}(V)$, while Spin double covers the special orthogonal group ${\rm SO}(V)$.

\subsection{Representation on polyforms}

There is a canonical realisation of the creation/annihilation operators satisfying (\ref{anti-comm}) in the space of differential forms (anti-symmetric tensors) $\Lambda \C^n$. Let us first review the construction, and then explain the geometry behind. 

We consider the space $\Lambda \mathbb{C}^n$ of mixed degree (that is inhomogeneous) anti-symmetric tensors (differential forms) on $\C^n$. We shall refer to such mixed degree forms as {\bf polyforms}. Let $e_i, i=1,\ldots, n$ be a basis in $\Lambda^1 \C^n$. We introduce the operators $a_i^\dagger$ of creation and $a_i$ of annihilation of basic 1-forms $e_i$. That is 
\be\label{cr-an}
a_i^\dagger \omega := e_i \wedge \omega, \qquad a_i \omega := e_i \lrcorner\, \omega,
\ee
where $\omega\in \Lambda\C^n$, $\wedge$ is the usual wedge product, and $ e_i \lrcorner$ is the operator that looks for an $e_i$ factor in $\omega$ and deletes it:
\be
e_i \lrcorner\, \left(e_{i_1} \wedge \ldots \wedge e_{i_k}\right)  =\sum_{m=1}^k  (-1)^{m-1} \delta_{i i_m} e_{i_1} \wedge \ldots ({\mathrm{omit\,\, mth\,\, factor}}) \ldots \wedge e_{i_k}.
\ee
It is a simple exercise to check that the introduced creation/annihilation operators satisfy the anti-commutation relations (\ref{anti-comm}), and so we get a concrete realisation of the Clifford algebra ${\rm Cl}(2n)$ by operators acting on $\Lambda \C^n$. Given that the Spin group sits inside the Clifford group, we get the realisation (\ref{spinors-polyforms}) of Dirac spinors of ${\rm Spin}(2n)$ as polyforms in $\Lambda \C^n$.

\subsection{Polyforms and complex structures}
\label{sec:polyforms-CS}

Consider the "empty" polyform, which is an element of $\Lambda^0 \C^n$. This spinor is annihilated by all the annihilation operators, and thus by $n$ Clifford generators
\be\label{subspace}
\Gamma_{i+n} + \frac{1}{\im} \Gamma_{i}.
\ee
Viewed as vectors in $\R^{2n}_\C$, they are null, and their pairwise inner products are zero. So, this set of vectors in $\R^{2n}_\C$ generates a maximal (dimension n) totally null subspace of $\R^{2n}_\C$. Such null subspaces arise naturally if one considers complex structures. 

\begin{definition}An orthogonal complex structure on $V$ is a map $J:V\to V$ that squares to minus the identity $J^2=-\id$, and that is metric compatible in the sense that $g(JX,JY)=g(X,Y)$. 
\end{definition}
We will omit the adjective orthogonal when no confusion can arise. Given a complex structure, the complexification $V_\C=V\otimes \C$ of $V$ splits into two subspaces, which are the eigenspaces of the operator $J$. Let us denote these subspaces by $E_\pm$, so that
\be
V_\C = E_+\oplus E_-.
\ee
It is a simple exercise to show that both $E_\pm$ are totally null, which means that the metric pairing of any two vectors in e.g. $E_+$ is zero. The non-zero metric pairings are those between $E_+$ and $E_-$. The real vectors in $V_\C$ are those whose projection to $E_+$ is the complex conjugate of the projection to $E_-$. 

A complex structure $J$ defines its eigenspaces $E_\pm$, but the reverse is also true. Thus, given a  totally null subspace $E_-\subset V_\C$ of dimension $n$, we can declare it to be the $-\im$ eigenspace of some complex structure $J$, and this characterises $J$ completely. Indeed, the knowledge of $E_-$ allows one to split any real vector $v\in V$ as $v=v_+ + v_-$ with $v_-\in E_-, v_+\in \overline{E_-}$. Then $Jv = \im (v_+ - v_-)$. 

Coming back to $V=\R^{2n}$ and the creation/annihilation operator model for ${\rm Cl}(2n)$, we have seen that this model comes with a preferred subspace of $\R^{2n}_\C$ spanned by (\ref{subspace}). Declaring thus null subspace to be $E_-$ defines a complex structure on $\R^{2n}$. This correspondence works in the opposite direction as well. One can start with a complex structure $J$ on $\R^{2n}$ and consider $\C^n=E_-$, the eigenspace of eigenvalue $-\im$ of $J$. One can then build the creation/annihilation operators as those acting on $\Lambda\C^n$. The totally null subspace of $\R^{2n}_\C$ that annihilates the empty polyform gives back $\C^n$ and thus the complex structure $J$.

Thus, the described model of ${\rm Cl}(2n)$ is in one-to-one correspondence with an orthogonal complex structure $J$ on $\R^{2n}$. We will rephrase this observation in terms of pure spinors in subsection \ref{sec:pure-spinors}.

\subsection{Spin Lie algebra, semi-spinors}

The (spinor representation of the) Lie algebra  $\mathfrak{spin}(2n)$ is generated by the commutators of $\Gamma$-matrices. Alternatively, given an anti-symmetric $2n\times 2n$ matrix $X^{AB}$ we can form the following operator acting on polyforms
\be\label{spin-2n}
A(X):= \frac{1}{2} \sum_{A<B} X^{AB} \Gamma_A \Gamma_B.
\ee
The map $A(X)$ introduced is the Lie algebra homomorphism in the sense that 
\be
[A(X), A(Y)]=A([X,Y]),
\ee
where on the right-hand side $[X,Y]$ is the commutator of two anti-symmetric matrices $X,Y$, i.e. $[X,Y]^{AB}=X^A{}_C Y^{CB}- Y^{A}{}_C X^{CB}$, and the index is lowered with the metric $\delta_{AB}$ on $\R^{2n}$.

The operators $A(X)$ act on the space $\Lambda(\C^n)$ of polyforms, and preserve the subspaces of even and odd degree polyforms. Thus, the space of polyforms split
\begin{equation}
    \Lambda \mathbb{C}^{n}=\Lambda^{\text{even}}\mathbb{C}^n\oplus\Lambda^{\text{odd}}\mathbb{C}^n,
\end{equation}
and each subspace is a representation of $\mathfrak{spin}(2n)$. We shall refer to the space of all polyforms as the space of (Dirac) {\bf spinors}. We will use the notation $S$ for this space, so our construction of the Clifford algebra identifies
\be
S=\Lambda \C^n.
\ee
The even and odd degree polyforms will be referred to as {\bf semi-spinors} (or Weyl spinors). We will call the even (odd) polyforms {\bf positive (negative)} semi-spinors. We will use the notation
\be
S^\pm = \Lambda^{even/odd} \C^n,
\ee
and $S=S_+ \oplus S_-$.

\subsection{Inner product}

The polyform description of spinors allows for a very simple description of the $\mathfrak{spin}(2n)$-invariant inner product on $S$. We call a polyform decomposable if it is homogeneous (i.e. of fixed degree) and can be written as the wedge product of one-forms. Let us introduce the operation $\sigma$ that rewrites each decomposable polyform in the opposite order
\be
\sigma( e_{i_1}\wedge e_{i_2} \wedge \ldots \wedge e_{i_k}) = e_{i_k} \wedge \ldots \wedge e_{i_2} \wedge e_{i_1}.
\ee
This operator extends to all of $\Lambda \C^n$ by linearity. 

Then, given two polyforms $\psi_1, \psi_2\in S$, the inner product is defined as
\be\label{inner-prod}
\langle \psi_1, \psi_2 \rangle = \sigma(\psi_1) \wedge \psi_2 \Big|_{top},
\ee
where the meaning of the right-hand side is that the wedge product of two polyforms is taken and then the projection to the top degree is applied. We remark that with this definition of the inner product any decomposable polyform is null.

The inner product is defined only after a top degree element from $\Lambda^n \C^n$ is chosen. To put it differently, an invariant inner product is only defined up to multiplication by a (complex-valued) constant. The invariance of the inner product defined under $\mathfrak{spin}(2n)$ transformations is an easy exercise using the following adjointness properties of the creation/annihilation operators
\be
\langle a_i \psi_1, \psi_2 \rangle=\langle \psi_1, a_i \psi_2 \rangle, \qquad \langle a_i^\dagger \psi_1, \psi_2 \rangle=\langle \psi_1, a_i^\dagger \psi_2 \rangle.
\ee

\subsection{Reality conditions, Majorana spinors}
\label{sec:RR'}

We now introduce two anti-linear maps $R, R'$ on $S$, which either commute or anti-commute with all $\Gamma$-matrices. As the result both $R, R'$ commute with all Lie algebra operators. Depending on $n$, these operators square to either plus or minus the identity operator. When we have an anti-linear operator that squares to the identity and commutes with all Lie algebra transformations, it is meaningful to restrict the action of $\mathfrak{spin}(2n)$ to one of the two eigenspaces of the anti-linear operator. This is how Majorana spinors arise.

We thus define the following anti-linear maps, given either by the product of all "real" $\Gamma$-matrices followed by the complex conjugation, or by the product of all "imaginary" $\Gamma$-matrices again followed by the complex conjugation
\begin{equation}
    R = \Gamma_{n+1} \ldots \Gamma_{2n} \ast, \qquad R' =  \Gamma_{1} \ldots \Gamma_{n} \ast.
\end{equation}
Here $\ast$ is the complex conjugation map, that is $\ast z = z^{*}$ for any $z \in \mathbb{C}^n$. The $\Gamma$-operators are then either real or imaginary 
\be
\ast \Gamma_{i+n} \ast = \Gamma_{i+n}, \qquad \ast \Gamma_{i}\ast = -\Gamma_{i}, \qquad i=1,\ldots, n.
\ee

We have the following lemma
\begin{lemma} 
The operators $R,R'$ either commute or anti-commute with all the $\Gamma$-operators
   \begin{equation}
    R \Gamma_A = (-1)^{n-1}\Gamma_A R, \qquad R' \Gamma_A = (-1)^{n}\Gamma_A R' \qquad A \in \{1, \ldots, 2n\}
    \end{equation} 
\end{lemma}
This means that both $R,R'$ are anti-linear operators that commute with all operators from $\mathfrak{spin}(2n)$. The proof is by verification. It can also be shown that, up to multiplication by a complex number, $R,R'$ are the only anti-linear operators that either commute or anti-commute with all Cliff$_{2n}$. A proof is analogous to the proof of Lemma 12.75 in \cite{Harvey}. 

We now need to establish a result on the square of each map
\begin{lemma}
    \begin{equation}\label{R2}
    R^2 = (-1)^{\frac{n(n-1)}{2}}\id  \qquad
    (R')^2 = (-1)^{\frac{n(n+1)}{2}}\id 
    \end{equation}
\end{lemma}
Proof is again a simple verification. 

We thus see that when $n$ is even $R^2=(R')^2$, and when $n$ is odd $R^2=-(R')^2$. Thus, when $n\in 4\mathbb{Z}$ both $R,R'$ square to plus the identity. When $n$ is even but not a multiple of 4, both $R,R'$ square to minus the identity, and neither gives a reality condition. When $n$ is odd either $R$ or $R'$ gives a reality condition. 

When there is a reality condition on $S$, i.e. an anti-linear operator $\mathcal R: {\mathcal R}^2=\id$ that commutes with all operators from $\mathfrak{spin}(2n)$, we can restrict the action of $\mathfrak{spin}(2n)$ to the {\bf Majorana} spinors, which is the subspace $S_M=\{ \psi \in S: {\mathcal R} \psi = \psi\}$. We have seen that there are no Majorana spinors when $n$ is even but not a multiple of four. 

We can also go a step further, and ask whether there exists a choice of the reality condition that can be imposed on the spaces of semi-spinors $S^\pm$. If such real semi-spinors exist they are called {\bf Majorana-Weyl spinors}. It is clear that only when $n$ is even both $R,R'$ are given by a product of an even number of $\Gamma$-matrices, and thus preserve $S^\pm$. However, we have seen that only when $n\in 4\mathbb{Z}$ we have a reality condition. 

The above discussion can be summarised as follows. When $n$ is odd, we have Majorana-Dirac spinors. When $n\in 4\mathbb{Z}$ we have Majorana-Weyl spinors. 

When $n$ is even $n\not\in 4\mathbb{Z}$ there are no Majorana spinors. But in this case there is an additional quaternionic structure. Indeed, both $R,R'$ square to minus the identity in this case, and in fact differ on $S_\pm$ by at most a sign, see below. Let us denote $J=R$. We can also consider the operator $I$ that multiplies all spinors by $\im$. It is clear that we have $J^2=I^2=-\id$, and that $IJ=-JI$, because $J$ is complex anti-linear. This gives the space of semi-spinors the quaternionic structure with 3 complex structures $I,J,K=IJ$. 

We also note that the products $RR', R'R$ coincide, modulo sign $(-1)^n$, with the product of all $\Gamma$-matrices
\be\label{RR-prime}
RR' = R'R = (-1)^n \Gamma_1 \ldots \Gamma_{2n}.
\ee
On the other hand, the product of all $\Gamma$-matrices squares to a multiple of the identity
\be
(\Gamma_1 \ldots \Gamma_{2n})^2= (-1)^n \id.
\ee
The eigenspaces of $\Gamma_1 \ldots \Gamma_{2n}$ are the spaces of semi-spinors $S_\pm$. When $n$ is even $R,R'$ preserve the spaces $S_\pm$ of semi-spinors, and the formula (\ref{RR-prime}) shows that these operators differ in their action on $S_\pm$ only by the sign, the eigenvalue of the operator $\Gamma_1 \ldots \Gamma_{2n}$. This means that, for $n$ even, there is (modulo sign) a unique anti-linear operator on semi-spinors, which can be taken to be either $R$ or $R'$. For $n$ odd the operators $R,R'$ do not preserve $S_\pm$, and the preferred anti-linear operator can always be taken the one that squares to the identity. 

\subsection{Distinguished ${\mathfrak u}(n)$ subalgebra}

Given that a creation/annihilation operator model for ${\rm Cl}(2n)$ relies on a choice of the complex structure on $\R^{2n}$, the Lie algebra $\mathfrak{spin}(2n)$ realised in terms of creation/annihilation operators comes with a distinguished ${\mathfrak u}(n)$ subalgebra. This is the subalgebra of $\mathfrak{spin}(2n)$ that does not mix the polyforms of different degrees in $\Lambda(\C^n)$. Its alternative characterisation is that it is generated by operators $a_i a_j^\dagger$ containing one creation and one annihilation operator. To see how this arises explicitly, let us rewrite the general Lie algebra element using indices of dimension $n$ rather than $2n$. We have
\be
A_{\mathfrak{spin}(2n)} = \frac{1}{2} X^{ij} (a_i + a_i^\dagger) (a_j + a_j^\dagger) - \frac{1}{2} \tilde{X}^{ij}  (a_i - a_i^\dagger) (a_j -a_j^\dagger) +\im  Y^{ij} (a_i + a_i^\dagger) (a_j - a_j^\dagger).
\ee
The matrices $X^{ij}, \tilde{X}^{ij}$ are anti-symmetric, while $Y^{ij}$ does not have any symmetry. The summation convention is implied. The subalgebra of this that does not contain the products $a_i a_j$ and $a_i^\dagger a_j^\dagger$ satisfies
\be
X^{ij} = \tilde{X}^{ij} , \qquad Y^{[ij]}=0.
\ee
Its general element is then
\be\label{un}
A_{{\mathfrak u}(n)} =  X^{ij} (a_i a_j^\dagger + a_i^\dagger a_j) - \im Y^{ij} (a_i a_j^\dagger - a_i^\dagger a_j) ,
\ee
which is anti-Hermitian. It is also not hard to check that the described ${\mathfrak u}(n)$ subalgebra is the one that in the vector representation acting on $\R^{2n}$ is compatible with the complex structure $J$ chosen.  

\subsection{Pure spinors and complex structures}
\label{sec:pure-spinors}

We can now rephrase the content of subsection \ref{sec:polyforms-CS} in terms of pure spinors.

The developed creation/annihilation operator model of ${\rm Cl}(2n)$ and $\mathfrak{spin}(2n)$ comes with a pair of preferred spinors. Indeed, we have the spinor given by the wedge product $e_1\wedge \ldots\wedge e_n$ of all $e_i$, which is the top degree polyform. Depending on the dimension, it is an element of either $S_+$ or $S_-$. The stabiliser of this spinor is $\mathfrak{su}(n)$. Another canonical spinor is the empty polyform. It is an element of $S_+$. Its stabiliser can be seen to be the same $\mathfrak{su}(n)$. Both the these spinors are examples of {\bf pure spinors}. 

If $\psi$ is a semi-spinor, denote by $\hat{\psi}$ the result of the action on $\psi$ of either of the anti-linear operators $R,R'$ discussed in subsection \ref{sec:RR'} 
\be\label{hat-spinor}
\hat{\psi}=R(\psi).
\ee
As discussed, when $n$ is even, the action of $R,R'$ coincide on $S_+$ and differ on $S_-$ by a sign. So, either of $R,R'$ can be used in (\ref{hat-spinor}), as the sign in the definition of $\hat{\psi}$ is immaterial to us. When $n$ is odd, the definition (\ref{hat-spinor}) should use that one of the operators $R,R'$ that squares to the identity. Note that $\hat{\psi}$ is of the same parity as $\psi$ when $n$ is even, and of the opposite parity when $n$ is odd. As is easy to check, the hat of the empty polyform is the top polyform and vice versa.

The spinor $e_1\wedge \ldots\wedge e_n$ is annihilated by all creation operators $a_i$, and so the dimension of the subspace of $\R^{2n}_\C$ that annihilates this spinor is $n$. Similarly, the empty polyform is annihilated by all the annihilation operators, and again the dimension of the annihilator subspace in $\R^{2n}_\C$ is $n$. This leads to the following
\begin{definition}
Let $\psi$ be a spinor. The vectors from $V$ act on $\psi$ by the Clifford multiplication. Denote by $V^0_\psi\subset V$ the subspace that annihilates $\psi$. The dimension of this subspace can be shown to be ${\rm dim}V^0_\psi \leq n$. If this dimension is maximal, i.e. ${\rm dim}(V^0_\psi)=n$, then $\psi$ is said to be a {\bf pure (or simple) spinor}. 
\end{definition}
It can be proved that the result of the action of ${\rm Spin}(2n)$ on a pure spinor is a pure spinor. It can also be shown that ${\rm Spin}(2n)$ acts transitively on the orbits of pure spinors of fixed length $\langle\hat{\psi},\psi\rangle$ in $S^\pm$, see e.g. \cite{Harvey} for a proof.

We have seen that choosing a complex structure $J$ on $\R^{2n}$ gives rise to the creation/annihilation operator model of ${\rm Cliff}_{2n}$, and to a preferred pure spinor $e_1\wedge \ldots\wedge e_n$ (or the empty polyform). As already discussed, this correspondence works in the opposite direction as well. Thus, each pure spinor $\psi$ defines a complex structure on $\R^{2n}$. Explicitly, this complex structure can be computed by first computing the real (or purely imaginary) 2-form
\be\label{moment-map}
M_{AB}:=\langle \hat{\psi}, \Gamma_A \Gamma_B \psi\rangle.
\ee
When $\psi$ is a pure spinor, the matrix $M_A{}^B$ with one of its indices raised with the metric on $\R^{2n}$ squares to a multiple of the identity, and its appropriate multiple is then the sought complex structure. 

\subsection{Example of ${\rm Cl}(4)$}

We now illustrate the above constructions on the example of Clifford algebra in four dimensions. 

We choose a complex structure thus identifying $\R^4=\C^2$. We will call the arising null complex coordinates $z_{1,2}$, and the corresponding one-forms $dz_{1,2}$. We introduce two pairs of creation/annihilation operators $a_{1,2}, a_{1,2}^\dagger$. The $\Gamma$ operators take the following form
\begin{equation}\label{gamma-spin4}
    \begin{split}
        \Gamma_4&=a_1+ a_1^{\dagger},\\
        \Gamma_2&=a_2 + a_2^{\dagger} ,\\
    \end{split}
    \qquad
    \begin{split}
        \Gamma_3&=i(a_1 - a_1^{\dagger} ),\\
        \Gamma_1&=i(a_2 - a_2^{\dagger} ).\\
    \end{split}
\end{equation}
 A generic Dirac spinor (general polyform) is given by 
\begin{equation}\label{psi-spin4}
    \Psi=(\alpha+\beta dz_{12})+(\gamma dz_1+\delta dz_2),
\end{equation}
where $dz_{12}:=dz_1\wedge dz_2$ and $\alpha,\beta,\gamma,\delta\in\C$. In matrix notations, the Dirac spinor is 4-component. It is convenient to adopt the $2\times 2$ block notations, in which Weyl spinors are 2-component. Thus, we write
\be
\Psi = \left(\begin{array}{c} \psi_+ \\ \psi_- \end{array}\right), \qquad
\psi_+ = \left(\begin{array}{c} \alpha \\ \beta \end{array}\right), \qquad
\psi_- = \left(\begin{array}{c} \gamma \\ \delta \end{array}\right).
\ee
The action of the $\Gamma$ operators is as follows
\be
\Gamma_4 \Psi = \alpha dz_1 + \beta dz_2 + \gamma+ \delta dz_{12}, \\ \nonumber
\Gamma_3 \Psi = -\im \alpha dz_1 + \im \beta dz_2 + \im \gamma- \im \delta dz_{12}, \\ \nonumber
\Gamma_2 \Psi =  - \beta dz_1 +\alpha dz_2+ \delta- \gamma dz_{12}, \\ \nonumber
\Gamma_1 \Psi = -\im \beta dz_1 - \im \alpha dz_2 + \im \delta+\im \gamma dz_{12}.
\ee
In matrix notations this becomes
\be\label{gamma-matr-spin4}
\Gamma_4 =\left( \begin{array}{cc} 0 & \id \\ \id & 0 \end{array}\right), \quad 
\Gamma_i = \left( \begin{array}{cc} 0 & \im \sigma^i \\ -\im \sigma^i & 0 \end{array}\right), \quad i=1,2,3.
\ee
Here $\sigma^i$ are the usual Pauli matrices. 

The Lie algebra is generated by products of distinct $\Gamma$-matrices. This gives a $4\times 4$ Lie algebra matrix that is block-diagonal. Let us refer to its $2\times 2$ blocks as $A, A'$, where $A$ acts on $S_+$ and $A'$ on $S_-$ respectively. We have
\be
A = \im (- \omega^{4i} + \frac{1}{2} \epsilon^{ijk} \omega^{jk} )\sigma^i , \qquad
A' = \im ( \omega^{4i} + \frac{1}{2} \epsilon^{ijk} \omega^{jk} )\sigma^i.
\ee
Both are anti-Hermitian $2\times 2$ matrices. This demonstrates $\mathfrak{spin}(4)=\mathfrak{su}(2)\oplus\mathfrak{su}(2)$.  

The invariant inner product is determined by the following computation
\be
\langle \Psi_1, \Psi_2\rangle = ( \alpha_1 - \beta_1 dz_{12} + \gamma_1 dz_1 + \delta_1 dz_2) \wedge ( \alpha_2 + \beta_2 dz_{12} + \gamma_2 dz_1 + \delta_2 dz_2) \Big|_{top} = \\ \nonumber
(\alpha_1 \beta_2-\alpha_2\beta_1) + (\gamma_1 \delta_2 - \gamma_2 \beta_1). 
\ee
It is thus an anti-symmetric pairing $\langle S_+, S_+\rangle, \langle S_-,S_-\rangle$. It can be written in matrix terms as
\be\label{inner-4}
\langle \Psi_1, \Psi_2\rangle = \Psi_1^T \left( \begin{array}{cc} \epsilon & 0 \\ 0 & \epsilon \end{array}\right) \Psi_2, \qquad \epsilon:= \im \sigma^2 =  \left( \begin{array}{cc} 0 & 1 \\ -1 & 0 \end{array}\right).
\ee
For the possible reality conditions, both $R=\Gamma_2 \Gamma_4 \ast$ and $R'=\Gamma_1 \Gamma_3\ast$ square to minus the identity, and so there are no Majorana spinors in this case. Of them $R'$ commutes with all $\Gamma$-matrices and defines the hat operator 
\be
\hat{\psi}_+ = \left(\begin{array}{c} \alpha \\ \beta \end{array}\right)^\wedge =\left(\begin{array}{c} -\beta^* \\ \alpha^* \end{array}\right) , \qquad
\hat{\psi}_-=\left(\begin{array}{c} \gamma \\ \delta \end{array}\right)^\wedge = \left(\begin{array}{c} \delta^* \\ -\gamma^* \end{array}\right),
\ee
which squares to minus the identity. 

There are two "canonical" pure spinors that come with the model, "identity" polyform $1$ and the top polyform $dz_{12}$. They are both in $S_+$. The stabiliser of both is the copy of $\mathfrak{su}(2)\subset \mathfrak{spin}(4)$ whose action on $S_+$ is trivial. 

It is clear that a generic Weyl spinor of ${\rm Spin}(4)$ is also pure, with stabiliser ${\rm SU}(2)$. The group ${\rm Spin}(4)$ acts transitively on the space of Weyl spinors of fixed norm $\langle \hat{\psi}_+, \psi_+\rangle$. This space is the 3-sphere $S^3$. 

As discussed, pure spinors are in one-to-one correspondence with complex structures. The complex structure on $\R^4$ corresponding to a generic Weyl spinor can be recovered as in (\ref{moment-map}). Let us consider the case of a spinor in $S_+$. A simple computation gives
\be\label{2-form-4D}
\langle \hat{\psi}_+, \Gamma\Gamma \psi_+\rangle =  \im \Sigma^i V^i, 
\ee
where
\be
\Sigma^i = dx^4 \wedge dx^i - \frac{1}{2} \epsilon^{ijk} dx^j \wedge dx^k
\ee
is the basis of self-dual 2-forms. The coordinates $x^{1,2,3,4}$ are those on $\R^4$. The vector $V^i$ is given by
\be
V^i := {\rm Tr}\left( \psi_+^\dagger \sigma^i \psi_+\right) = ( 2 {\rm Re}(\alpha^*\beta),   2 {\rm Im}(\alpha^*\beta), |\alpha|^2-|\beta|^2),
\ee
and is a 3-vector with squared norm 
\be
|V|^2=V^i V^i = (|\alpha|^2+|\beta|^2)^2 = \langle \hat{\psi}_+, \psi_+\rangle^2.
\ee
If one raises an index of $\Sigma^i_{\mu\nu}$, one obtains a triple of endomorphisms of $\R^{4}$ that satisfy the algebra of the quaternions
\be
\Sigma^i_{\mu}{}^{\rho} \Sigma^j_\rho{}^\nu = - \delta^{ij} \delta_\mu{}^\nu + \epsilon^{ijk} \Sigma^k_\mu{}^\nu.
\ee
The object
\be
J_{\psi_+} := \frac{1}{|V|} \Sigma^i V^i,
\ee
viewed as an endomorphism of $\R^4$, is then a complex structure that corresponds to the pure spinor $\psi_+$. 

\section{Octonionic models of ${\rm Cl}(-6)$, ${\rm Cl}(8)$ and ${\rm Cl}(10)$}
\label{sec:oct}

As discussed in subsection \ref{sec:RR'}, there exists an anti-linear operator on the spaces of semi-spinors $S_\pm$ of ${\rm Spin}(8)$ that squares to the identity. Thus, there exist real semi-spinors (Majorana-Weyl spinors) of ${\rm Spin}(8)$. The space of such positive and negative parity spinors is real 8-dimensional. Given two spinors of opposite parity $\phi_+\in S_+, \phi_-\in S_-$ we can form the vector $\langle \phi_+, \Gamma \phi_-\rangle\in \R^8$. This gives a map $\R^8\times \R^8\to \R^8$. This map is the octonionic product, and the existence of octonions can thus be traced to the existence of Majorana-Weyl spinors of ${\rm Spin}(8)$. Majorana-Weyl spinors of ${\rm Spin}(8)$ become identified with octonions, and this observation implies that there is an octonionic model for ${\rm Cl}(8)$ and ${\rm Spin}(8)$. The Clifford algebra ${\rm Cl}(-6)$ sits inside ${\rm Cl}(8)$, and thus also admits an octonionic description. The Clifford algebra ${\rm Cl}(10)$ contains ${\rm Cl}(8)$ and can also be described using octonions. 

\subsection{Creation/annihilation operator construction}

While the material described in this section is known, it is hard to point out a reference containing all the details. We worked out the details in a recent work \cite{Bhoja:2022txw}, from which we take the constructions described in this and several subsections that follow.

We start by spelling out the details of the creation/annihilation operator construction of ${\rm Cl}(8)$. As we have explained above, this Clifford algebra knows about the octonions, and the algebra of octonions can be derived rather than postulated. 

As in the case of ${\rm Cl}(4)$, we start by choosing a complex structure. This identifies $\R^8\sim \C^4$. We denote the complex coordinates by $z_{1,\ldots,4}$, and the coordinate one-forms by $dz_{1,\ldots,4}$. We introduce four pairs of creation/annihilation operators $a^\dagger_I, a_I, I=1,\ldots, 4$. We introduce the following $\Gamma$-operators
\be\label{gamma-spin8}
\Gamma_{4+I} := a_I+ a_I^{\dagger} , \qquad \Gamma_{I} := \im (a_I- a_I^{\dagger} ), \quad I=1,2,3,4.
\ee
Spinors are polyforms, i.e. elements of $\Lambda(\C^4)$, in general with complex coefficients. Weyl spinors are even or odd degree polyforms. 

The relevant anti-linear operator that arises in this case is given by the product of all four imaginary $\Gamma$-operators with the complex conjugation
\be
R := \Gamma^1 \Gamma^2 \Gamma^3 \Gamma^4 *,
\ee
The operator $R$ commutes with all the gamma-matrices and $R^2=\id$, so $R$ is a real structure. Since $R$ is composed of an even number of gamma-matrices, it preserves the spaces of Weyl spinors. Thus, it allows us to define the notion of Majorana-Weyl spinors for ${\rm Spin}(8)$. A simple computation shows that the action of $R$ is complex conjugation followed by the Hodge star in $\C^4$, modulo some signs. 
 
 \subsection{Majorana-Weyl spinors explicitly}
 
A general odd/even polyform that is also real (in the sense of describing a Majorana-Weyl spinor) can be written as follows
 \be\label{real-polyforms}
 \psi^- \equiv \psi^-(u)= u_1 dz_11 + \bar{u}_1 dz_{234} + u_2 dz_2 + \bar{u}_2 dz_{314}+ u_3 dz_3 +\bar{u}_3 dz_{124} +\im \bar{u}_4 dz_4 +  \im u_4 dz_{123}, \\ \nonumber
 \psi^+ \equiv \psi^+(u) = 
  u_1 dz_{41}+\bar{u}_1 dz_{23} +u_2 dz_{42}+\bar{u}_2 dz_{31} +u_3 dz_{43}+\bar{u}_3 dz_{12} +\im \bar{u}_4+\im u_4 dz_{4123}  ,
 \ee
 where the quantities $u_I$, with $I=1,\ldots,4$ are complex numbers and $dz_{I\ldots J}= dz_I\wedge\ldots \wedge dz_J$. This particular choice of the complex coordinates $u_I$, and in particular the choices made for the coordinate $e^4$, will become justified below by the desired form of the action of the $\Gamma$-matrices.

The inner product is a pairing $\langle S_+, S_+\rangle, \langle S_-, S_-\rangle$. If we take two positive spinors $\psi^+(\tilde{u}), \psi^+(u)$, the product $\langle \psi^+(\tilde{u}), \psi^+(u)\rangle$ is computed by taking the polyform $\psi^+(\tilde{u})$ in the reverse order, wedging with $\psi^+(u)$ and projecting on the top component. A simple computation gives
\be
\langle \psi^+(\tilde{u}), \psi^+(u)\rangle= 2{\rm Re}\sum_{I=1}^4  \tilde{u}_{I} \bar{u}_{I}.
\ee
Thus, Majorana-Weyl spinors are identified $S_\pm \sim \C^4$, with the invariant pairings on $S_\pm$ being given by the standard definite Hermitian metric on $\C^4$. 

The form of the inner product makes it clear that the basis $dz_{42}, dz_{42}, dz_{43}, 1$ of $S^+$ in (\ref{real-polyforms}) is totally null. To make contact with octonions that are non-null objects, it is necessary to switch to a different parametrisation of polyforms. We parametrise the polyforms by the real and imaginary parts of $u_I$ and write
\be\label{u-ab}
u_1= \alpha_1+\im  \alpha_5, \quad u_2 =  \alpha_2 + \im  \alpha_6, \quad u_3=  \alpha_3+\im  \alpha_7, \quad u_4= \alpha_0 + \im  \alpha_4.
\ee
We will denote the components of the positive polyform by $\alpha$ and the negative polyform components by $\beta$. We then have
\be\label{polyforms-octonions}
\psi^+ = \alpha_1( dz_{41}+dz_{23}) + \alpha_2( dz_{42}+dz_{31}) +\alpha_3( dz_{43}+dz_{12}) + \alpha_4 (1- dz_{4123}) \\
\nonumber
+ \im \alpha_5( dz_{41}-dz_{23}) + \im \alpha_6( dz_{42}-dz_{31}) +\im \alpha_7( dz_{43}-dz_{12}) + \im \alpha_0 (1+ dz_{4123}) ,
\\ \label{polyforms-octonions-negative}
\psi^- = \beta_1( dz_{1}+dz_{423}) + \beta_2( dz_{2}+dz_{431}) +\beta_3( dz_{3}+dz_{412}) + \beta_4 (dz_4- dz_{123}) \\
\nonumber
+ \im \beta_5( dz_{1}-dz_{423}) + \im \beta_6( dz_{2}-dz_{431}) +\im \beta_7( dz_{3}-dz_{412}) + \im \beta_0 (dz_4+ dz_{123}).
\ee

 \subsection{The action of $\Gamma$-matrices}
 
 We now introduce a 16-component column 
 \be
\Psi = \left( \begin{array}{c} \psi^+ \\ \psi^-\end{array}\right) = \left( \begin{array}{c} \alpha_0 \\ \alpha_1 \\ \vdots \\ \alpha_7 \\ \beta_0 \\ \beta_1 \\ \vdots \\ \beta_7  \end{array}\right).
\ee
A computation shows that the $\Gamma$-operators become the following $16\times 16$ matrices
\begin{equation}\label{Gamma-matr-spin8}
    \Gamma_0=
    \begin{pmatrix}
    0&\id \\
    \id &0
    \end{pmatrix}
    , \quad
    \Gamma_a=
    \begin{pmatrix}
    0&-L_a\\
    L_a&0
    \end{pmatrix}
    \ \text{for} \ a\in \{1,\ldots,7\},
\end{equation}
where
\begin{equation}\label{E-matr-spin8}
    \begin{split} 
        L_1&=-E_{01}+E_{27}-E_{36}+E_{45},\\
        L_2&=-E_{02}-E_{17}+E_{35}+E_{46},\\
        L_3&=-E_{03}+E_{16}-E_{25}+E_{47},\\
        L_4&=-E_{04}-E_{15}-E_{26}-E_{37},\\
        L_5&=-E_{05}+E_{14}+E_{23}-E_{67},\\
        L_6&=-E_{06}-E_{13}+E_{24}+E_{57},\\
        L_7&=-E_{07}+E_{12}+E_{34}-E_{56}.
    \end{split}
\end{equation}
Here $E_{ij}$ are the standard generators of $\mathfrak{so}(8)$. Each is an anti-symmetric matrix whose only non-zero entries are in the $i$-th column and $j$-th row, and in the $j$-th column and $i$-th row. The $i$-th column $j$-th row element is $+1$. It can be checked explicitly that these matrices square to minus the identity $L_a^2=-\id$, are anti-symmetric, and anti-commute with each other. At the same time, all these properties are guaranteed because matrices $\Gamma_I$ satisfy the defining relations of ${\rm Cl}(8)$ by construction. 

Note that all the $\Gamma$-matrices, in the parametrisation of spinors by the real quantities $\alpha_0,\ldots,\alpha_7, \beta_0,\ldots,\beta_7$, are real. This is as expected because the $\Gamma$-matrices commute with the reality condition operator, and are therefore real. 

The octonions are manifest already at this stage, because the matrices $L_a$ can be interpreted as the operators of multiplication by the imaginary octonions. To see this we form the matrix
\be
L_x : = x_0 \id + \sum_{a=1}^7 x_a L_a.
\ee
It is then a simple check to see that
\be
L_x \left(\begin{array}{c} 1\\0 \\ 0\\0 \\0 \\0 \\0 \\0 \end{array}\right) = \left(\begin{array}{c} x_0\\x_1 \\ x_2\\x_3 \\x_4 \\x_5 \\x_6 \\x_7 \end{array}\right).
\ee
The interpretation is that 8-component columns are identified with octonions, the column on the left is the identity octonion, and the operator $L_x$ is the operator of left multiplication by the octonion $x$. We develop this octonionic interpretation in more details in the next subsection.

\subsection{Octonions}

There are several different ways to arrive to the octonions. The popular account in \cite{Baez:2001dm} presents them as the result of the Cayley–Dickson construction. In the previous subsection we have seen octonions arising from the description of the Clifford algebra ${\rm Cl}(8)$. Let us summarise their properties. 

Octonions are objects that can be represented as linear combinations of the unit octonions $\textbf{1} ,{\bf e}^1,\ldots, {\bf e}^7$
\be
x = x_0 \textbf{1} + x_1 {\bf e}^1 +\ldots + x_7 {\bf e}^7, \qquad x_0, x_1, \ldots, x_7 \in \R.
\ee
The conjugation is the operation that flips the signs of all the imaginary coefficients
\be
\overline{x} = x_0 \textbf{1} - x_1 {\bf e}^1 -\ldots - x_7 {\bf e}^7,
\ee
and we have 
\be
|x|^2=x\overline{x}= (x_0)^2 + (x_1)^2 + \ldots + (x_7)^2 \equiv |x|^2.
\ee
If we represent an octonion as an 8-component column with entries $x_0,x_1,\ldots, x_7$, denoted by the same symbol $x$, we can write the norm as $|x|^2\equiv(x,x)=x^T x$. 

Octonions $\Oc$ form a normed division algebra that satisfies the highly non-trivial composition property $|xy|^2=|x|^2|y|^2$. The only non-trivial part of the octonionic product is that of two imaginary octonions. The product of two imaginary octonions is encoded in the matrices (\ref{E-matr-spin8}). To recover it we just need to take the product of two matrices $L_a L_b, a\not=b$. This can then be projected onto $L_c$ to obtain the structure constants of the imaginary octonion multiplication. Explicitly, we form three imaginary octonions $x=x_a L_a, y=y_a L_a, z=z_a L_a, x,y,z\in{\rm Im}(\Oc)$, and consider
\be\label{C-product}
(x \textbf{1}, y z \textbf{1}) = -(\textbf{1}, xyz \textbf{1}) = C_{abc} x_a y_b z_c.
\ee
Here $\textbf{1}$ is the 8-component column $\textbf{1} = (1,0,0,0,0,0,0,0)$, and we have defined
\be
C_{abc} := - ( \textbf{1}, L_a L_b L_c \textbf{1} ).
\ee
The object $C$ is an anti-symmetric rank 3 tensor $C\in \Lambda^3(\R^7)$. An explicit computation, which is best carried by a symbolic manipulation software such as Mathematica, gives
\be\label{C-R7}
C = e^{567} + e^5\wedge (e^{41}-e^{23}) + e^6\wedge (e^{42}-e^{31}) + e^7\wedge (e^{43}-e^{12}),
\ee
where the notation is $e^{ijk}=e^i\wedge e^j\wedge e^k$. We see from (\ref{C-product}) that the interpretation of $C_{abc} x_a y_b z_c$ is that this is the left action of the imaginary octonion $y$ on the imaginary octonion $z$, projected onto the octonion $x$. In other words, the tensor $C_{abc}$ encodes the table of the products of the imaginary octonions
\be
C(x,y,z)=(x, yz), \qquad x,y,z\in {\rm Im}(\Oc).
\ee
For example, ${\bf e}^5 {\bf e}^6={\bf e}^7$, which is captured by the first term in (\ref{C-R7}). 

Octonions are non-commutative and non-associative, but alternative. The last property is equivalent to saying that any two imaginary octonions (as well as the identity) generate a subalgebra that is associative, and is a copy of the quaternion algebra $\Hq$. For more details on the octonions, see e.g. \cite{Karigiannis2}.

\subsection{The octonionic model of ${\rm Cl}(8)$}

As we have discussed, the $L_a$ matrices in (\ref{Gamma-matr-spin8}) are given the interpretation of the operators of left multiplication by a unit imaginary octonion, and the 8-component columns they act on are identified with octonions. The formulas in (\ref{Gamma-matr-spin8}) thus give the following octonionic model of the Clifford algebra ${\rm Cl}(8)$
\be\label{oct-model-cl8}
\Gamma_x = \left( \begin{array}{cc} 0 & L_{\bar{x}} \\ L_x & 0 \end{array}\right).
\ee
Here we have introduced an octonion $x$ with components $x_I, I=1,\ldots, 8$, and defined the linear combination of $\Gamma$-matrices $\Gamma_x:= \sum_I x_I \Gamma_I$. The operators $L_x, L_{\bar{x}}$ are the operators of left multiplication by an octonion $x$ and its conjugate $\bar{x}$. The Clifford generators act on 2-component columns (Dirac spinors) with entries in $\Oc$, which identifies (real) semi-spinors $S_\pm= \Oc$. 

\subsection{Automorphisms of the octonions and some of its subgroups}

The cross-product of the imaginary octonions, encoded by the 3-form $C$ (\ref{C-R7}), is preserved by $G_2\subset{\rm SO}(7)$, the smallest rank exceptional Lie group. This is the group of automorphisms of the octonions (and imaginary octonions). 

The maximal subgroups of $G_2$ are ${\rm SU}(3)$ and ${\rm SO}(4)={\rm SU}(2)\times{\rm SU}(2)/\Z_2$. The group ${\rm SU}(3)$, and the form of the octonionic product that exhibits this subgroup will be discussed in the next subsection. The group ${\rm SO}(4)$ is manifest already in (\ref{C-R7}). 

Indeed, the 3-form $C$ is written in a way that exhibits the $3+4$ split of the imaginary octonions. The three directions are those spanned by $5,6,7$, and the four directions are $1,2,3,4$. To put it differently, the 3-form (\ref{C-R7}) views the imaginary octonions as ${\rm Im}(\Oc)=\Hq\oplus{\rm Im}(\Hq)$, that is as a copy of quaternions spanning the directions $1,2,3,4$ and the imaginary quaternions spanning $5,6,7$. The objects
\be
\Sigma^1 := e^{41}-e^{23}, \qquad \Sigma^2:=e^{42}-e^{31}, \qquad \Sigma^3:=e^{43}-e^{12}
\ee
are the basis of self-dual (or anti-self-dual, depending on the orientation) 2-forms in $\R^4$. The group of rotations in $\R^4$ ${\rm SO}(4)={\rm SU}(2)\times{\rm SU}(2)/\Z_2$ acts on this triple of 2-forms in a very simple way. They are left invariant by one of the two ${\rm SU}(2)$ factors, while the other ${\rm SU}(2)$ mixes them as in the spin one representation of ${\rm SU}(2)$. The expression (\ref{C-R7}) makes is clearly invariant under the simultaneous rotation of the $\Sigma^{1,2,3}$ and the factors $e^{5,6,7}$. This exhibits the ${\rm SO}(4)$ symmetry of (\ref{C-R7}).

\begin{remark} There is another common choice of the multiplication rules for the octonions, the one exhibiting the index cycling ${\bf e}_{i+1} {\bf e}_{j+1}={\bf e}_{k+1}$, and index doubling properties ${\bf e}_{2i} {\bf e}_{2j}={\bf e}_{2k}$, all mod 7. This choice is adopted e.g. in \cite{Baez:2001dm}. The 3-form $C$ encoding the products in this case is given by
\be
C_{\rm Baez}= e^{124}+e^{137}+e^{165}+e^{235}+e^{267}+e^{346}+e^{457}.
\ee
This choice of the multiplication table does not exhibit the ${\rm Im}(\Oc)=\Hq\oplus{\rm Im}(\Hq)$ structure and makes the geometry more obscure. Our choice is standard in the $G_2$ geometry literature, see e.g. \cite{Karigiannis1}, \cite{Karigiannis2} and \cite{Agricola:2014yma}, with the exception that we choose the imaginary octonionic units $5,6,7$ rather than $1,2,3$ to lie in ${\rm Im}(\Hq)$. See also Remark 4 below for a further discussion of this point. 
\end{remark}

\subsection{Clifford algebra ${\rm Cl}(-6)$ and ${\rm SU}(3)$ geometry}
\label{sec:cl6}

The Clifford algebras ${\rm Cl}(-6),{\rm Cl}(-7)$ are isomorphic to ${\rm End}(\R^8)$ and ${\rm End}(\R^8)\oplus{\rm End}(\R^8)$ respectively. Both can be recovered from our model (\ref{oct-model-cl8}) for ${\rm Cl}(8)$. Indeed, taking the Clifford generators $\Gamma_a, a=1,\ldots, 7$, we notice that the operators $L_x, x\in{\rm Im}(\Oc)$ generate ${\rm End}(\R^8)$, and operators $-L_x, x\in{\rm Im}(\Oc)$ generate another copy of ${\rm End}(\R^8)$. 

We take ${\rm Cl}(-6)$ to be generated by $L_{1,2,3,5,6,7}$, which are the operators of left multiplication by the unit imaginary octonions ${\bf e}^{1,2,3,5,6,7}$ respectively. As operators acting on $\R^8$ they are given by the matrices $L_{1,2,3,5,6,7}$ and moreover we have
\be\label{product-six}
L_1 L_2 L_3 L_5 L_6 L_7 = L_4
\ee
Thus, the product of all 6 $\Gamma$-matrices of ${\rm Cl}(-6)$ squares to minus the identity and is a complex structure on the spaces $S_\pm$ of Weyl spinors of ${\rm Cl}(-7)$.
\begin{remark} As explained previously, our way of treating the octonions exhibits the ${\rm Im}(\Oc)=\Hq\oplus{\rm Im}(\Hq)$ structure, where the last three octonionic imaginary units are in ${\rm Im}(\Hq)$ and the first four are in $\Hq$. The choice to have the last three imaginary octonionic units in ${\rm Im}(\Hq)$ rather than the first three is not completely standard, as most the of $G_2$ literature uses conventions that ${\bf e}^{1,2,3}\in{\rm Im}(\Hq)$. Our choice is motivated by the geometry of the Hopf fibration $S^7\to S^4$. With the conventions ${\bf e}^{5,6,7}\in{\rm Im}(\Hq)$ these directions become those along the fibres, while ${\bf e}^{1,2,3,4}\in\Hq$ parametrise the base $S^4$. This choice is based on the intuition that $S^4$ is the Euclidean version of "our" four-dimensional world, and the fibre directions are "extra dimensions".  After the choice of which units are assigned to ${\rm Im}(\Hq)$ is made, to describe ${\rm Cl}(-6)$ one needs to make the further choice as to which of the four octonionic units is identified with the identity in $\Hq$. The choice we make is to select ${\bf e}^4$ as special in this sense. 
\end{remark}

From the general discussion of the available anti-linear operators on Clifford algebras in subsection \ref{sec:RR'} we can conclude that there exists an anti-linear operator that commutes with all the ${\rm Cl}(-6)$ generators and squares to plus the identity. To see this, one needs to extend the creation/annihilation operator construction to Clifford algebras ${\rm Cl}(-n)$. This is done by placing the imaginary units in the analog of (\ref{Gamma-a}) in front of the sum of the creation and annihilation operators rather than the difference. The anti-linear operator one gets for ${\rm Cl}(-6)$ maps $S_+\to S_-$ and vice versa. It is thus meaningful to impose the Majorana-Dirac condition. A Weyl spinor of ${\rm Cl}(-6)$ is an even or odd polyform in $\Lambda(\C^3)$, and the space of such polyforms is complex 4-dimensional. The Majorana-Dirac spinor of ${\rm Cl}(-6)$ is one where the semi-spinor in $S_-$ is the complex conjugate of the semi-spinor in $S_+$. Thus, a Majorana-Dirac spinor of ${\rm Cl}(-6)$ is an object in $\R^8$, and this explains why a model of ${\rm Cl}(-6)$ in terms of real $8\times 8$ matrices is possible. 

The action of the complex structure $L_4$ on $S_+=\Oc$ decomposes the complexification of this space into the two eigenspaces of $L_4$:
\be
S_+^\C = \Oc^\C = E_+\oplus E_-,
\ee
and the explicit calculation using the octonion multiplication table shows that 
\be
E_- = {\rm Span}({\bf e}^1+\im {\bf e}^5,  {\bf e}^2+\im {\bf e}^6,  {\bf e}^3+\im {\bf e}^7,  \textbf{1}+ \im {\bf u}).
\ee
We have denoted ${\bf u}\equiv{\bf e}^4$. It is interesting to translate these (complex) spinors into polyforms using (\ref{polyforms-octonions}). If these complexified octonions are interpreted as elements of $S_+$, the space of even polyforms, we see that
\be\label{octonions-polyforms-even}
{\bf e}^1+\im {\bf e}^5 \leftrightarrow dz_{23}, \qquad {\bf e}^2+\im {\bf e}^6 \leftrightarrow dz_{31} , \qquad {\bf e}^3+\im {\bf e}^7 \leftrightarrow dz_{12} , \qquad \textbf{1}+ \im {\bf u} \leftrightarrow \im 1.
\ee
In the last expression $\textbf{1}$ is the identity octonion, and $1$ is the "empty" polyform $1\in\Lambda^0(\C^4)$. We observe that the spinors that are in the $-\im$ eigenspace of $L_4$ are in fact (even) polyforms in $\Lambda(\C^3)$, precisely those even polyforms in $\Lambda(\C^4)$ that do not involve the $dz_4$ direction:
\be \label{L4-minus}
E_- = \Lambda^{even}(\C^3).
\ee
Similarly, it is not hard to see that the $+\im$ eigenspace of $L_4$ in $S_+^\C$ is spanned by $dz_4$ wedged with the odd polyforms in $\Lambda(\C^3)$:
\be \label{L4-plus}
E_+ = \Lambda^{odd}(\C^3).
\ee
A real semi-spinor of ${\rm Spin}(8)$ then splits into two complex conjugate semi-spinors of ${\rm Spin}(6)$, or in other words a Majorana-Dirac spinor splits into two complex conjugate Weyl spinors. Given that ${\rm Spin}(6)$ commutes with the complex structure on $\R^8$ given by $L_4$, it admits a complex description ${\rm Spin}(6)={\rm SU}(4)$. 

Let us also state the same correspondence but for odd polyforms. From (\ref{polyforms-octonions}), the analogous translation of the complexified octonions into odd polyforms is
\be\label{octonions-polyforms-odd}
{\bf e}^1+\im {\bf e}^5 \leftrightarrow dz_{423}, \qquad {\bf e}^2+\im {\bf e}^6\leftrightarrow dz_{431}, \qquad {\bf e}^3+\im {\bf e}^7\leftrightarrow dz_{412}, \qquad \textbf{1}+ \im {\bf u}\leftrightarrow \im dz_{4}.
\ee
This is the product of even polyforms in $\Lambda(\C^3)$ with $dz_4$. Similarly, the complexified octonions that are eigenvectors of $L_4$ with the other sign correspond to odd polyforms in $\Lambda(\C^3)$. Thus, the isomorphisms (\ref{L4-minus}), (\ref{L4-plus}) also hold for $S_-^\C$. 

Any semi-spinor $\phi$ of ${\rm Spin}(6)$ is a pure spinor. This will become obvious from the general facts on pure spinors to be reviewed in the next Section. Thus any semi-spinor determines a complex structure on $\R^6$. Let us describe this explicitly. We follow \cite{Agricola:2014yma}. 

Let $\phi\in S_+\sim \C^4$ of ${\rm Spin}(6)$. Then, viewed as a real vector space $S_+$ splits into three components
\be\label{s-plus-decomp}
S_+ = \R \phi \oplus \R L_4(\phi) \oplus \{ X\cdot \phi: X\in \R^6\}.
\ee
Here $\cdot$ denotes the Clifford multiplication. In particular, $L_4$ preserves the subspace $\{ X\cdot \phi: X\in \R^6\}$. For example, when $\phi=\textbf{1}$ the identity octonion, the three factors become $\R \textbf{1}\oplus \R {\bf u} \oplus {\rm Span}({\bf e}^{1,2,3,5,6,7})$. The complex structure $J_\phi$ on $\R^6$ corresponding to a spinor $\phi$ is then given by
\be
J_\phi(X) \cdot \phi = L_4( X\cdot \phi).
\ee
For the spinor $\phi=\textbf{1}$ this is the complex structure $J_{\phi=\textbf{1}}(X) = L_4 X$, where $\R^6$ is identified with the six directions $1,2,3,5,6,7$. 

A spinor $\phi\in S_+$ defines not only a complex structure on $\R^6$ but also a 3-form (that is moreover the real or imaginary part of the top holomorphic form for the complex structure $J_\phi$). The 3-form is given by
\be
\Omega_\phi (X,Y,Z) := -(\phi, X\cdot Y\cdot Z\cdot \phi). 
\ee
For the spinor $\phi=\textbf{1}$ this 3-form is given by
\be
\Omega\equiv \Omega_{\phi=\textbf{1}} = - {\rm Im}( (e^1+\im e^5)(e^2+\im e^6) (e^3+\im e^7)),
\ee
so that the 3-form (\ref{C-R7}) that encodes the product of imaginary octonions can be written as
\be\label{C-su3}
C = e^4 \wedge \omega + \Omega, \qquad \omega:= e^1\wedge e^5+ e^2\wedge e^6 + e^3\wedge e^7,
\ee
and $\Omega$ is just the restriction of $C\in\Lambda^3(\R^7)$  to $\R^6$, while $\omega$ is the K\"ahler 2-form for $J_{\phi=\textbf{1}}$. Recalling that the 3-form $C$ encodes the cross-product product of imaginary octonions, the form (\ref{C-su3}) of writing this 3-form exhibits the ${\rm SU}(3)$ subgroup of the group of automorphisms of the octonions $G_2$. 

All in all, any semi-spinor $\phi\in S_+$ of ${\rm Spin}(6)$ is pure and gives rise to a complex structure on $\R^6$, together with the top holomorphic form, thus breaking ${\rm Spin}(6)={\rm SU}(4)$ to ${\rm SU}(3)$. The space of spinors of fixed length is $S^7$, and so the space of complex structures (together with a holomorphic top form) on $\R^6$ is 
\be
S^7 = {\rm SU}(4)/{\rm SU}(3).
\ee
The group ${\rm Spin}(6)$ acts on the space of spinors of fixed length $(\phi,\phi)$ in $S_+$ transitively, and any unit spinor can be mapped to the identity octonion spinor by this action. 

\begin{remark} The decomposition (\ref{s-plus-decomp}) is the decomposition $\C^4=\C\oplus \C^3$ that played the key role in the constructions in \cite{Dubois-Violette:2016kzx}-\cite{Todorov:2019hlc}. We thus see how such a decomposition of Majorana-Dirac spinors $S_+=\R^8=\C^4$ of ${\rm Spin}(6)$ is parametrised by a choice of a unit such spinor, which breaks ${\rm Spin}(6)$ to ${\rm SU}(3)$. 
\end{remark}

\subsection{Clifford algebra ${\rm Cl}(10)$}

We now apply the constructions of Section \ref{sec:2n} to the case of ${\rm Cl}(10)$, taking into account the developed above octonionic description of ${\rm Cl}(8)$. The outcome is an octonionic model for ${\rm Cl}(10)$. It will be useful to relate the arising this way model to the polyform description of Section \ref{sec:2n}, rather than just state the model, because this clarifies some geometry.

Choosing a complex structure in $\R^{10}$, the spinors of ${\rm Cl}(10)$ become polyforms in $\Lambda(\C^5)$. In the previous subsections we have seen that real polyforms in $\Lambda(\C^4)$ can be identified with octonions. This means that general complex polyforms in $\Lambda(\C^4)$ can be identified with complexified octonions. This means that we can write even and odd polyforms in $\Lambda(\C^4)$ as 
\be\label{psi-pm-10}
\psi_+ = \alpha + \beta \wedge dz_5, \qquad \psi_- = \gamma  \wedge dz_5 +\delta,
\ee
where $\alpha,\gamma \in \Lambda^+(\C^4)$ and $\beta,\delta\in \Lambda^-(\C^4)$, all identified with complexified octonions. From (\ref{oct-model-cl8}) we know that the generators $\Gamma_I, I=1,\ldots,8$ act on $\Lambda^+(\C^4)$ as $L_x$ and on $\Lambda^-(\C^4)$ as $L_{\bar{x}}$. This means that if we arrange all the components into a Dirac spinor 
\be\label{spin10-Dirac}
\left(\begin{array}{c} \alpha \\ \beta \\ \gamma \\ \delta \end{array}\right) , \qquad \alpha,\beta,\gamma,\delta\in\Oc_\C
\ee
we have the following $32\times 32$ matrices for the generators $\Gamma_I, I=1,\ldots,8$
\be\label{gamma-x}
\Gamma^x = \left( \begin{array}{cccc} 0 & 0 & 0 & L_{\bar{x}} \\ 0 & 0 & L_x & 0 \\ 0 & L_{\bar{x}} & 0 & 0 \\ L_x & 0 & 0 & 0 \end{array}\right).
\ee
The generators $\Gamma_{9,10}$ are then easily computed using the creation/annihilation operator construction 
\be
\Gamma_9 = a_5^\dagger + a_5, \qquad \Gamma_{10} = \im (a_5^\dagger - a_5).
\ee 
An easy computation that takes into account that creation/annihilation operators commute with even polyforms in $\Lambda^+(\C^4)$ (and thus octonions $\alpha,\gamma$) and anti-commute with odd-polyforms (and thus $\beta,\delta$) gives
\be\label{gamma-910}
\Gamma^9 = \left( \begin{array}{cccc} 0 & 0 & 1 & 0 \\ 0 & 0 & 0 & -1 \\ 1 & 0 & 0 & 0 \\ 0 & -1 & 0 & 0 \end{array}\right), \qquad 
\Gamma^{10} = \left( \begin{array}{cc} 0 & -\im \id  \\ \im \id & 0 \end{array}\right).
\ee
Of all the $\Gamma$-matrices only the matrix $\Gamma_{10}$ is complex (purely imaginary). Thus, spinors of ${\rm Cl}(10)$ are necessarily complex, and our model identifies Dirac spinors with 4-component columns (\ref{spin10-Dirac}) with values in $\Oc_\C$ and Weyl spinors with 2-component such columns. 

\subsection{Lie algebra ${\mathfrak so}(10)$}

The commutators of the ${\rm CL}(10)$ $\Gamma$-matrices are block-diagonal. This means that the Lie algebra $\mathfrak{so}(10)$, in one of its two semi-spinor representations, can be represented by $2\times 2$ matrices with complexified octonionic entries. Indeed, the top diagonal $16\times 16$ block can rewritten as a $2\times 2$ matrix with entries in complexified octonions
\be\label{spin-10} 
\left( \begin{array}{cc} A+ \im a & - L_{\bar{x}}+\im L_{\bar{y}} \\ L_x +\im L_y  & A'-\im a \end{array}\right), \qquad A,A' \in \mathfrak{so}(8), \quad x,y\in \Oc, \quad a\in \R.
\ee
This gives an explicit description of the Lie algebra ${\mathfrak so}(10)$ in terms of $2\times 2$ matrices with complexified octonionic entries, acting on 2-columns with entries in complexified octonions. This gives the octonionic model of the $\mathfrak{so}(10)$ algebra. This model is similar to the one that appears in \cite{Bryant}, but the route we obtained it here is different from the one taken in this reference.

\subsection{Spinor inner product, charge conjugation}

As we have seen above, the semi-spinor representations of ${\rm Spin}(10)$ are $S_\pm = \Oc^2_\C$. There is a ${\rm Spin}(10)$ invariant pairing between $S_+$ and $S_-$. It is not difficult to describe this pairing explicitly, starting from the general expression (\ref{inner-prod}). Denoting the semi-spinors by 
\be\label{sp}
 S_+ \ni \psi_+=\left( \begin{array}{c} \alpha \\ \beta \end{array}\right), \qquad 
 S_- \ni \psi_-=\left( \begin{array}{c} \gamma \\ \delta \end{array}\right), 
 \ee
the ${\rm Spin}(10)$ invariant inner product between them is given by 
\be\label{inner-prod-spin10}
\langle \psi_+, \psi_-\rangle = (\alpha, \gamma) + (\beta, \delta),
\ee
where $(\cdot,\cdot)$ is the octonionic product. Note that the inner product is in general complex. 

The general discussion in subsection \ref{sec:RR'} tells us that there exists the ${\rm Spin}(10)$-invariant anti-linear operator $R$ that commutes with all $\Gamma$ matrices and squares to plus the identity. This operator is given by the product of all five "real" $\Gamma$-operators, followed by the complex conjugation. Thus, this operator maps $S_+\to S_-$, and vice versa. Given that the product of four real $\Gamma$-operators followed by the complex conjugation acts on complexified octonions as simply the complex conjugation, it is an easy calculation to check that the ${\rm Spin}(10)$ $R$-operator is given by 
\be
S_+\ni \psi_+ = \left( \begin{array}{c} \alpha_1 + \im \alpha_2  \\ \beta_1 + \im \beta_2  \end{array}\right) \to R(\psi_+) = \left( \begin{array}{c} \alpha_1 - \im \alpha_2  \\ \beta_1 - \im \beta_2  \end{array}\right) \in S_-.
\ee
With this map in place we have a ${\rm Spin}(10)$ invariant Hermitian bilinear form on $S_+$
\be
\langle R(\psi_+), \psi_+ \rangle = |\alpha_1|^2+|\alpha_2|^2+|\beta_1|^2 + |\beta_2|^2. 
\ee

\section{Pure spinors}
\label{sec:pure}

The main aim of this Section is to describe some general facts about pure spinors. Most of these facts were known already to Eli Cartan, see \cite{Cartan}. Another classical reference is \cite{Chevalley}. 

Pure spinors for ${\rm Spin}(2n)$ are Weyl spinors. In dimensions six and lower all spinors are pure, as will be easily seen from Cartan's purity criterion below. The first dimension where one encounters impure spinors is thus eight. Majorana-Weyl spinors in dimension eight, which as we already discussed are identified with octonions, are impure spinors. Thus, octonions are the simplest impure spinors that exist. The fact that the geometry of octonions is interesting and non-trivial thus suggests that the geometry of impure spinors should be interesting as well. While the geometry of pure spinors is well-understood, as we have reviewed above, the study of the geometry of impure spinors is only taking its first steps. 

Equipped with understanding of the pure spinor geometry in dimensions eight and ten, we describe the possible types of impure (i.e. general) spinors that can arise in these dimensions. As we shall see, the behaviour of impure spinors in these two dimensions is the same, which makes it natural to discuss them together.

\subsection{Pure spinors and MTN}

As before, let $V$ be a real vector space of dimension ${\rm dim}(V)=2n$, equipped with a definite metric. We denote its complexification $V_\C=W$. Given a spinor $\psi$ we define $M(\psi):= W^0_\psi$, the subspace of the complexification $W$ of $V$ that annihilates $\psi$ via the Clifford product. The spinor $\psi$ is said to be pure if the dimension of the space $M(\psi)$ is maximal possible, i.e. $n$. We shall refer to $M(\psi)$ as the {\bf maximal totally null} (MTN) subspace corresponding to the pure spinor $\psi$. It is known since Cartan \cite{Cartan} that pure spinors are Weyl. Cartan also gives a very useful algebraic characterisation of pure spinors, which we now explain.

Given a (Weyl) spinor $\psi$ (or a pair of Weyl spinors $\psi,\phi$), one can insert a number of $\Gamma$-matrices between two copies of $\psi$ (or, more generally, between $\psi$ and $\phi$) Let us introduce the convenient notation
\be
\Lambda^k(\R^{2n})\ni B_k(\psi,\phi) := \langle \psi, \underbrace{\Gamma\ldots \Gamma}_{\text{$k$ times}} \phi\rangle.
\ee
It is assumed that distinct $\Gamma$-matrices are inserted, thus giving components of a degree $k$ differential form (anti-symmetric tensor) in $\R^{2n}$. 

The following proposition is due to Cartan \cite{Cartan} and Chevalley \cite{Chevalley}
\begin{theorem} \label{thm:cartan} A Weyl spinor $\phi$ is a pure (or in Cartan's terminology simple) if and only if $B_k(\phi,\phi) =0$ for $k\not=n$. The $n$-vector $B_n(\phi,\phi)$ is then proportional to the wedge product of the vectors constituting a basis of $M(\phi)$, where $M(\phi)$ is the MTN subspace that corresponds to $\phi$. 
\end{theorem}
We note that this theorem gives a practical way of recovering $M(\phi)$ as the set of vectors whose insertion into $B_n(\phi,\phi)$ vanishes. Thus, this theorem establishes a one-to-one correspondence between pure spinors $\phi$ (up to rescaling) and MTN subspaces $M(\phi)$. This theorem also gives a set of quadratic constraints that each pure spinor must satisfy. 

\subsection{Pure spinors in dimensions up to and including eight}
\label{sec:pure8}

Weyl spinors in dimensions four and six are all pure. One can easily see this from Cartan's criterion. 

Let us start in dimension four. Let $\phi\in S_+$. The inner product is an anti-symmetric pairing $\langle S_+,S_+\rangle$, which means that $\langle \phi,\phi\rangle=0$. Thus, $B_0(\phi,\phi)=0$ and the spinor is pure. The 2-form $B_2(\phi,\phi)$ is decomposable and its factors are the null directions spanning $M(\phi)$. 

In dimension six the inner product is a pairing $\langle S_+, S_-\rangle$. One thus needs to insert an odd number of $\Gamma$ matrices between two copies of a Weyl spinor $\phi\in S_+$. The objects $B_k(\phi,\phi)$ that can be constructed are thus $B_1(\phi,\phi)$ and $B_3(\phi,\phi)$. However, the $\Gamma$-matrices of ${\rm Cl}(-6)$ are anti-symmetric and $B_1(\phi,\phi)=0$. For example, this can be seen in the octonionic model for ${\rm Cl}(-6)$ described in subsection (\ref{sec:cl6}). Then only $B_3(\phi,\phi)$ is different from zero. It is automatically decomposable, with the factors spanning the MTN corresponding to $\phi$. 

In dimension eight the inner product $\langle S_+,S_+\rangle$ is symmetric, which means that there exist spinors $\phi\in S_+$ for which $B_0(\phi,\phi)\not=0$. Thus, there exist impure spinors in dimension eight. In fact, for Majorana-Weyl spinors that we identified with octonions the inner product $\langle\phi,\phi\rangle$ is just the norm squared of the corresponding octonion. It is never zero for a non-zero $\phi$. Thus, Majorana-Weyl spinors of ${\rm Cl}(8)$ are impure. Pure spinors are null, and thus necessarily complexified octonions. For example, the spinor $\phi=\textbf{1}+\im {\bf u}\in \Oc_\C$ is null. To conclude that this spinor is pure we also need to consider $B_2(\phi,\phi)$. However, the product of two $\Gamma$-matrices is anti-symmetric and hence  $B_2(\phi,\phi)=0$ for any spinor. This means that the only constraint that a spinor must satisfy to be pure in this dimension is to be null. 

The complex structure corresponding to a pure spinor $\phi\in S_+$ of ${\rm Cl}(8)$ can be described explicitly using the octonionic model. Let $\phi= \alpha_1 + \im \alpha_2, \alpha_{1,2}\in \Oc$ be a complexified octonion. Then 
\be
B_0(\phi,\phi)= (\alpha_1+\im\alpha_2,\alpha_1+\im\alpha_2) = |\alpha_1|^2-|\alpha_2|^2+ 2\im (\alpha_1,\alpha_2). 
\ee
The spinor is then pure if $|\alpha_1|^2=|\alpha_2|^2, (\alpha_1,\alpha_2)=0$. Given a null complexified octonion we can compute the corresponding complex structure explicitly. The general prescription (\ref{moment-map}) is to compute the 2-form $B_2(R(\phi),\phi)$. When one of the two indices of this 2-form are raised, it becomes an endomorphism of $\R^8$ that squares to a multiple of the identity, and so the complex structure sought is an appropriate rescaling of this endomorphism. Using the octonionic model (\ref{oct-model-cl8}) we get
\be
\langle R(\phi), \Gamma_x \Gamma_y \phi\rangle = (\alpha_1-\im\alpha_2, L_{\bar{x}} L_y (\alpha_1+\im\alpha_2)) = \im (\alpha_1, (L_{\bar{x}} L_y - L_{\bar{y}} L_x) \alpha_2).
\ee
The resulting 2-form is thus purely imaginary. Dividing by (twice) the imaginary unit and raising one of the indices with the flat metric we get the following endomorphism of $\R^8$
\be\label{end-R8}
M(x)= \frac{1}{2} (x \alpha_2) \bar{\alpha}_1 - \frac{1}{2} (x \alpha_1) \bar{\alpha}_2.
\ee
We can simplify this using the identity
\be\label{oct-identity}
(a b) \bar{c} + (ac)\bar{b}= 2(b,c) a.
\ee
Assuming now that $\alpha_1+\im\alpha_2$ is null and using (\ref{oct-identity}) we see that both terms in (\ref{end-R8}) are equal. The square of the endomorphism (\ref{end-R8}) is then easily seen to be
$M^2= -|\alpha_1|^2|\alpha_2|^2 \id= - |\alpha_1|^4 \id$, and so 
\be\label{J-R8}
J_\phi(x) = \frac{(x \alpha_2)\bar{\alpha}_1}{|\alpha_1|^2} 
\ee
is the complex structure that corresponds to the pure (null) spinor $\phi=\alpha_1+\im\alpha_2\in S_+$. For later purposes we note that the direction $\bar{\alpha}_1-\im \bar{\alpha}_2\in \R^8_\C$ is the eigenvector of $J_\phi$ of eigenvalue $-\im$. As an example of this complex structure, for the pure spinor $\phi=\textbf{1}+\im {\bf u}$ the complex structure is $J_{\textbf{1}+\im {\bf u}}= R_{\bf u}$, the operator of right multiplication by the unit imaginary octonion ${\bf u}$. 

In the next subsection we will need also the complex structure as defined by a pure spinor of the other parity. Thus, let $\psi=\beta_1+\im\beta_2 \in S_-, |\beta_1|^2=|\beta_2|^2, (\beta_1,\beta_2)=0$ be a pure negative spinor. A similar computation gives 
\be
\langle R(\psi), \Gamma_x \Gamma_y \psi\rangle = (\beta_1-\im\beta_2, L_x L_{\bar{y}} (\beta_1+\im\beta_2)) = \im (\beta_1, (L_x L_{\bar{y}} - L_y L_{\bar{x}}) \beta_2),
\ee
and then
\be\label{J-negative-R8}
J_\psi(x) = \frac{\beta_1(\bar{\beta}_2 x)}{|\beta_1|^2}
\ee
is the corresponding complex structure on $\R^8$. In contrast to (\ref{J-R8}) it is the left multiplication by octonions $\alpha_{1,2}$ that is now involved. We note, for later purposes, that $\beta_1-\im \beta_2$ is an eigenvector of $J_\psi$ of eigenvalue $-\im$. 

It is clear that (\ref{J-R8}) is invariant under the rescaling of $\alpha_1+\im\alpha_2$ by an arbitrary complex number, as expected. Let us count the number of parameters needed to parametrise $J$. It is parametrised by two vectors in $\R^8$ of equal length that are orthogonal to each other. Moreover, by rescaling these vectors can be be made unit. Thus, $J$ is parametrised by a point in the total space of the unit tangent bundle over $S^7$. This gives a 13-dimensional configuration space. One still needs to mod out by the freedom of multiplying $\alpha_1+\im\alpha_2$ by the phase $e^{\im\theta}$. This leaves a 12-dimensional configuration space parametrising $J$. 

On the other hand, the space of complex structures on $\R^8$ is 
\be\label{Z8}
Z_8 = {\rm SO}(8)/{\rm U}(4),
\ee
which is ${\rm dim}(Z_8)= 28 - 16 = 12$ dimensional. This matches the previous count as it should. Another way to arrive at the same dimension is to note that the space of complex structures is the space of pure spinors modulo rescaling. The space of pure spinors of ${\rm Cl}(8)$ is the null quadric in $\C^8$, and taken projectively is the quadric in $\mathbb{CP}^7$. The quadric in $\mathbb{CP}^7$ is complex six-dimensional, or real 12-dimensional, which matches the dimension of (\ref{Z8}). The space $Z_8$ can be described as the total space of a fibre bundle over $S^6$, with fibres $\mathbb{CP}^3$, see e.g. \cite{Salamon}.

\subsection{Pure spinors in ten dimensions}
\label{sec:pure10}

Let us now consider the case of pure spinors in ten dimensions. There is a known explicit parametrisation of pure spinors in this dimension obtained by restricting ${\rm Spin}(10)$ to a ${\rm U}(5)$ subgroup, after which the components of a spinor are parametrised by $0,2,4$ forms in $\C^5$. The purity condition can be solved explicitly in the patch where the 0-form part is non-vanishing, see e.g. \cite{Nekrasov:2005wg}, formula (3.40). In this subsection we derive an alternative octonionic parametrisation.  

The inner product is a pairing $\langle S_+,S_-\rangle$, and so an odd number of $\Gamma$-matrices can be inserted between two copies of an $S_+$ spinor. The object $B_3(\phi,\phi)$ is automatically zero due to anti-symmetry of the corresponding pairing. So, the only possible objects that can be constructed from two copies of the same positive parity spinor are $B_1(\phi,\phi)$ and $B_5(\phi,\phi)$. Cartan's criterion then tells us that a spinor $\phi\in S_+$ is pure if and only if $B_1(\phi,\phi)=0$.

We now parametrise $\phi\in S_+$ by a pair of complexified octonions $\alpha,\beta\in \Oc_\C$. A computation using the explicit form (\ref{gamma-x}), (\ref{gamma-910}) of the $\Gamma$-matrices gives
\be
\langle \phi, \Gamma \phi\rangle = X+ \im Y, \qquad X,Y\in \R^{10},
\ee
where
\be\label{comput-XY}
X= ( 2(\alpha_1, L_{\bar{x}} \beta_1) - 2(\alpha_2, L_{\bar{x}} \beta_2), |\alpha_1|^2-|\alpha_2|^2 -|\beta_1|^2 + |\beta_2|^2, -2 (\alpha_1,\alpha_2) - 2(\beta_1,\beta_2)), \\ \nonumber
Y= ( 2(\alpha_1, L_{\bar{x}} \beta_2) + 2(\alpha_2, L_{\bar{x}} \beta_1), 2 (\alpha_1,\alpha_2) - 2(\beta_1,\beta_2) ,|\alpha_1|^2-|\alpha_2|^2 +|\beta_1|^2 -|\beta_2|^2).
\ee
Here $x$ is the placeholder for the octonionic directions in $\R^{10}$, and the last two components are the 9th and 10th ones. An explicit computation shows that 
\be\label{X2-Y2}
|X|^2+|Y|^2= 2\langle R(\phi),\phi\rangle^2 - 8 I(\phi), \qquad |X|^2=|Y|^2, \qquad (X,Y)=0,
\ee
where $I(\phi)$ is the quartic invariant, see \cite{Bryant}
\be\label{quartic}
I(\phi) =|\alpha_1|^2|\alpha_2|^2  + |\beta_1|^2 |\beta_2|^2-(\alpha_1,\alpha_2)^2 - (\beta_1,\beta_2)^2+(\beta_1 \bar{\alpha}_1, \beta_2 \bar{\alpha}_2) -(\beta_1 \bar{\alpha}_2, \beta_2 \bar{\alpha}_1) .
\ee
So, in general a Weyl spinor of ${\rm Spin}(10)$ gives rise to a complex null vector $X+\im Y$, whose squared norm is related to the two invariants one has for a Weyl spinor of ${\rm Spin}(10)$. Note that for a pure spinor with both $X,Y$ vanishing, the formula (\ref{X2-Y2}) shows that the quartic invariant is a multiple of the square of the quadratic invariant, and so there is just one independent invariant in the case of ${\rm Spin}(10)$ pure spinors. 

For a pure spinor both $X,Y$ must vanish, which gives the following conditions
\be\label{su5-equations}
\beta_1 \bar{\alpha}_1 - \beta_2 \bar{\alpha}_2=0, \quad \beta_1 \bar{\alpha}_2 + \beta_2 \bar{\alpha}_1=0, \quad (\alpha_1,\alpha_2)= (\beta_1,\beta_2)=0, \quad |\alpha_1|^2=|\alpha_2|^2, \quad |\beta_1|^2=|\beta_2|^2. 
\ee
The last four equations are already familiar to use from the analysis in the previous subsection. They imply that $\alpha_1+\im\alpha_2$ and $\beta_1+\im\beta_2$ are pure spinors of ${\rm Cl}(8)$. Thus, each defines a complex structure on $\R^8$. These spinors are of different parities, and so $\alpha_1+\im\alpha_2$ gives rise to the complex structure (\ref{J-R8}), and $\beta_1+\im\beta_2$ gives (\ref{J-negative-R8}). 

To analyse the implications of the first two equations in (\ref{su5-equations}) we make the assumption that $\alpha_1\not=0$. We then multiply the first and second equation by $\alpha_1$ from the right. We get
\be\label{su5-temp1}
\beta_1 |\alpha_1|^2 - (\beta_2 \bar{\alpha}_2) \alpha_1 =0, \qquad (\beta_1 \bar{\alpha}_2) \alpha_1 + \beta_2 |\alpha_1|^2=0.
\ee
The second of these gives 
\be\label{beta2-beta1}
\beta_2 = - \frac{ (\beta_1 \bar{\alpha}_2)\alpha_1}{|\alpha_1|^2}.
\ee
One can then check that the first of the equations in (\ref{su5-temp1}) is automatically satisfied when (\ref{beta2-beta1}) holds together with the equations $|\alpha_1|^2=|\alpha_2|^2, (\alpha_1,\alpha_2)=0$. Further, it is easy to check that $|\beta_1|^2=|\beta_2|^2, (\beta_1,\beta_2)=0$ are automatic when (\ref{beta2-beta1}) holds. 

Thus, under the assumption that $\alpha_1\not=0$, all equations in (\ref{su5-equations}) are solved by a pair of orthogonal octonions of the same norm $|\alpha_1|^2=|\alpha_2|^2, (\alpha_1, \alpha_2)=0$, and another octonion $\beta_1$ that is arbitrary. The first pair of octonions forms a pure spinor $\alpha_1+\im \alpha_2$ of ${\rm Cl}(8)$, and the complexified octonion $\beta_1+\im \beta_2$, with $\beta_2$ given by (\ref{beta2-beta1}) is also a pure spinor of ${\rm Cl}(8)$. Let us verify that the dimension count is correct. A pure spinor of ${\rm Spin}(8)$ is a null vector in $\C^8$, and so complex dimension seven. A general octonion $\beta_1$ has real dimension eight, complex dimension four. Together we have $7+4=11$, which is the correct complex dimension of pure spinors in ten dimensions. The obtained explicit solution to the purity constraints can be viewed as an octonionic alternative to the parametrisation (3.40) in \cite{Nekrasov:2005wg}. 

The parametrisation of the ${\rm Spin}(10)$ pure spinor we obtained is not new, and has been previously obtained and used in the pure spinor superstring formalism, see e.g. \cite{Berkovits:2001mx}, formula (2.1), where a pure spinor is parametrised by a pure spinor of ${\rm Spin}(8)$ and a vector in $\R^8$. This is precisely our parametrisation by the ${\rm Spin}(8)$ pure spinor given by the null octonion $\alpha_1+\im \alpha_2$ and a real octonion $\beta_1$. The novelty of our discussion here is the octonionic parametrisation that is used. 

Another possible approach to ${\rm Spin}(10)$ pure spinors is to restrict to the ${\rm Spin}(6)={\rm SU}(4)$ subgroup. A Weyl spinor of ${\rm Spin}(10)$ then becomes a collection of four spinors of ${\rm SU}(4)$, and the pure spinor constraint can be solved in this parametrisation as well. For details, see \cite{Grassi:2004xc}.

The space of pure spinors of ${\rm Spin}(10)$ can be viewed as the complex cone over the space 
\be
Z_{10} = {\rm SO}(10)/{\rm U}(5)
\ee
of orthogonal complex structures in $\R^{10}$. This complex cone has complex dimension 11, and can be shown to admit a Calabi-Yau metric. In particular, there is a holomorphic top form. This geometry  is discussed in \cite{Cederwall:2011yp}.

\subsection{Intersections of pure spinors}
\label{sec:inters}

To understand the geometric implication of (\ref{beta2-beta1}), and also to understand impure spinors in dimensions eight and ten, we need to address the question of how the complex structures defined by pure spinors may be compatible. This is related to the question of whether the sum of two pure spinors can be pure. There is a beautiful geometry that is related to this question, which we now explain.

Let us start with $\phi,\psi$ being two pure spinors of ${\rm Cl}(2n)$ of the same parity, with the corresponding MTN subspaces $M(\phi), M(\psi)$. We have the following set of statements, most due to Cartan \cite{Cartan}
\begin{theorem}\label{thm:same} The MTN subspaces of two pure spinors of the same parity $\phi,\psi\in S_+$ can intersect along $n-2m, m\in\mathbb{N}$ null directions. The intersection along $n-2m$ directions occurs if and only if $B_k(\phi,\psi)=0$ for $k<n-2m$ and $B_{n-2m}(\phi,\psi)$ is non-zero. It is then decomposable and  its factors are the common null directions in $M(\phi), M(\psi)$. Two MTN's $M(\phi), M(\psi)$ do not intersect if and only if $B_0(\phi,\psi)\not=0$. 
\end{theorem}
For a proof, see e.g. \cite{Budinich:1989bg}, Proposition 9. We also have the following fact due to Chevalley
\begin{theorem} When the MTN subspaces of two pure spinors $\phi,\psi$ intersect along $n-2$ directions, the sum $\phi+\psi$ is a pure spinor. 
\end{theorem}
This last fact follows from the fact that all Weyl spinors in four dimensions are pure. Its proof can also be found in \cite{Budinich:1989bg}.

Let us now consider the case of $\phi,\psi$ being two pure spinors of ${\rm Cl}(2n)$ of opposite parity. The following statement is again due to Cartan
\begin{theorem}\label{thm:opposite} The MTN subspaces of two pure spinors of opposite parity $\phi\in S_+, \psi\in S_-$ can intersect along $n-2m-1, m\in\mathbb{N}$ null directions. The intersection along $n-2m-1$ directions occurs if and only if $B_k(\phi,\psi)=0$ for $k<n-2m-1$ and $B_{n-2m-1}(\phi,\psi)$ is non-zero. It is then decomposable and  its factors are the common null directions in $M(\phi), M(\psi)$. Two MTN's $M(\phi), M(\psi)$ do not intersect if and only if $B_0(\phi,\psi)\not=0$. 
\end{theorem}

\subsection{Intersection of the null subspaces of octonions giving a pure spinor of ${\rm Spin}(10)$}

Theorem \ref{thm:opposite} allows us to discuss the geometric meaning of the compatibility condition (\ref{beta2-beta1}) between the two null octonions that make up a pure spinor of ${\rm Spin}(10)$. We have two pure spinors $\phi=\alpha_1+\im\alpha_2$ and $\psi=\beta_1+\im \beta_2$ of ${\rm Cl}(8)$ of opposite parities. The positive parity spinor defines the complex structure (\ref{J-R8}), while the negative parity spinor defines the complex structure (\ref{J-negative-R8}). It is clear that (\ref{beta2-beta1}) says something about the compatibility of these two complex structures. According to Theorem \ref{thm:opposite}, the MTN subspaces intersect in 3 or 1 null directions. A moment of reflection shows that the first two conditions in (\ref{su5-equations}) that require the vanishing of the first eight components of $X,Y$ just encode the condition $B_1(\phi,\psi)=0$. This means that the MTN subspaces $M(\phi), M(\psi)$ intersect in 3 null directions. 

The single null directions in $M(\phi), M(\psi)$ that do not align can be computed as follows. The vanishing (for a pure spinor) null direction in $\R^8_\C$ was computed in (\ref{comput-XY}) as a multiple of $\langle\phi,L_{\bar{x}} \psi\rangle$. Both $\phi,\psi$ are pure spinors (of opposite parity) with four null directions, and the vanishing of $\langle\phi,L_{\bar{x}} \psi\rangle$ implies that their null subspaces intersect in three null directions. This means that the null directions of $R(\phi)=\alpha_1-\im \alpha_2$, which are complementary to those of $\phi$, must intersect with the null directions of $\psi$ in one dimension. This direction can be computed as $\langle R(\phi),L_{\bar{x}} \psi\rangle$, and is given by
\be
\langle R(\phi),L_{\bar{x}} \psi\rangle = (\alpha_1, L_{\bar{x}} \beta_1) + (\alpha_2, L_{\bar{x}} \beta_2)+ \im (\alpha_1, L_{\bar{x}} \beta_2)-\im (\alpha_2, L_{\bar{x}} \beta_1).
\ee
However, when $\langle\phi,L_{\bar{x}} \psi\rangle$ vanishes, the two real and the two imaginary terms are the same, and so we have
\be
\langle R(\phi),L_{\bar{x}} \psi\rangle = 2 (\alpha_1, L_{\bar{x}} \beta_1) -2\im (\alpha_2, L_{\bar{x}} \beta_1) = 
2( x, \beta_1(\bar{\alpha}_1- \im \bar{\alpha}_2)).
\ee
We can instead collect the terms containing $\beta_2$ and write the same quantity as
\be
\langle R(\phi),L_{\bar{x}} \psi\rangle = 2 (\alpha_2, L_{\bar{x}} \beta_2) +2\im (\alpha_1, L_{\bar{x}} \beta_2) = 
2\im ( x, \beta_2(\bar{\alpha}_1- \im \bar{\alpha}_2)).
\ee
This computation shows that the complex (null) vector that can be written in two alternative ways
\be\label{xi}
\xi:= \beta_1(\bar{\alpha}_1- \im \bar{\alpha}_2)= \im \beta_2(\bar{\alpha}_1- \im \bar{\alpha}_2)
\ee
should be the single null direction that is not shared by the null spaces of $\phi,\psi$. This can be confirmed by an explicit calculation. Indeed, we have compute
\be
J_\phi(\xi) = \frac{((\beta_1(\bar{\alpha}_1- \im \bar{\alpha}_2))\alpha_2)\bar{\alpha}_1}{|\alpha_1|^2}= -\im \beta_1(\bar{\alpha}_1- \im \bar{\alpha}_2) = -\im \xi,
\ee
where we have used the facts $|\alpha_1|^2=|\alpha_2|^2, (\alpha_1,\alpha_2)=0$, and the identity (\ref{oct-identity}). Similarly, 
\be
J_\psi(\xi) = \frac{\beta_1 (\bar{\beta}_2(\im \beta_2(\bar{\alpha}_1- \im \bar{\alpha}_2)))}{|\beta_1|^2} = \im \beta_1 (\bar{\alpha}_1- \im \bar{\alpha}_2) = \im \xi.
\ee
Thus, the null vector $\xi$ is in the $-\im$ eigenspace of $J_\phi$, while in the $+\im$ eigenspace of $J_\psi$. 

Let us also note that the null vector $\xi$ can be written more compactly as half the sum of two different versions (\ref{xi}) of writing it. Indeed, we can write
\be\label{xi*}
2\xi = (\beta_1+\im \beta_2)(\bar{\alpha}_1-\im \bar{\alpha}_2) = \psi \bar{\phi}^*.
\ee
This way of writing is of course not surprising, as this is precisely the way we extracted $\xi$ from $\langle R(\phi),L_{\bar{x}} \psi\rangle$. The vector $\xi$ is thus just the product of the two null octonions, the one that represents the spinor $\psi$, and the complex conjugation of the bar conjugation of the octonion representing the spinor $\phi$. 

\subsection{Product structures}

We want to prove that the compatibility conditions (\ref{beta2-beta1}) between the null octonions $\phi,\psi$ are equivalent to the statement that the two complex structures $J_\phi, J_\psi$ commute. This is stated as Proposition \ref{pure-10} below. To prove it, we need to remind some geometric structures arising when one has two commuting complex structures. 
  
  \begin{definition} A product structure on a vector space $V$ (equipped with a positive definite metric $g(\cdot,\cdot)$) is an endomorphism $K:V\to V$ such that $K^2=+\id$ and $g(Kx,Ky)=g(x,y), x,y\in V$.
  \end{definition}
  \begin{remark} The second condition $g(KX,Ky)=g(X,Y)$ is a compatibility condition between an endomorphism $K$ that squares to plus the identity and the metric. For this reason it is possible to refer to $K:V\to V$ such that $K^2=+\id$ as a product structure, and to the one that satisfies all the conditions of the above definition as an orthogonal product structure. In this paper we will always mean orthogonal product structure when we say product structure.
  \end{remark}
  
  \begin{proposition} The eigenspaces of $K$ of eigenvalue $\pm 1$ are metric orthogonal and so $V=V_+\oplus V_-$. Thus $K$ splits $V$ into a sum of two orthogonal subspaces. This justifies the name "product structure".
  \end{proposition}
  To prove this we consider two vectors $x\in V_+, y\in V_-$ from different eigenspaces. Thus $Kx=x, Ky=-y$. Therefore we have
  \be
  g(x,y)=-g(Kx,Ky)= - g(x,y),
  \ee
  which implies that this inner product is zero. 
  
  \subsection{Product structures compatible with complex structures}
  \label{sec:prod-struct}
  
  We now consider the situation when we are given both a complex structure on $V=\R^{2n}$ and a product structure. 
  \begin{definition} A product structure $K$ on $V$ is called J-compatible with a complex structure $J$ on $V$ if they commute $JK=KJ$.
  \end{definition}
 We then have the following 
  \begin{proposition} A product structure on $V=\R^{2n}$ that is J-compatible with a complex structure $J$ on $V$ provides a decomposition $\C^n=\C^k\oplus\C^l$. The two subspaces $\C^k,\C^l\subset\C^n$ are the two different eigenspaces of $K$.
  \end{proposition}
 To see that this statement holds we note that when $J,K$ commute they can be simultaneously diagonalised. We have previously denoted the eigenspaces of $J$ by $E_\pm\subset V_\C$. They are totally null. The map $K$ can be extended to $V_\C$ by linearity. Because $J,K$ commute $K$ applied to $E_\pm$ preserves this spaces. This means that $E_+=\C^n$ splits into the $\pm 1$ eigenspaces of the operator $K$
 \be
 E_+ = E_+ \cap V^\C_+ \oplus E_+ \cap V^\C_-,
 \ee
 where $V^\C_\pm$ are the eigenspaces of $K$ on $V_\C$. This is the promised $\C^n = \C^k\oplus \C^l$ split. 
  
  We now note that the product operator $JK$, when $J,K$ commute, is another complex structure
  \begin{proposition} Let $J$ be a complex structure on $V$, and $K$ be a J-compatible product structure. Then $JK$ is another orthogonal complex structure on $V$, i.e. $(JK)^2=-\id, g(JKx,JKy)=g(x,y)$.
  \end{proposition}
  Both properties follow in an elementary way from those of $J,K$ and the fact that the two commute. We note that the original complex structure $J$ commutes with the new complex structure $JK$. 
  
  Thus, when we have a complex structure and a compatible with it product structure, there is a second complex structure that this pair defines. The reverse is also true. 
  \begin{proposition} Given a pair of commuting complex structures $J_1,J_2$, their product $K:=J_1 J_2$ is a product structure. 
  \end{proposition} 
  Thus, a complex structure and a compatible with it product structure are equivalent to a pair of commuting complex structures. 
  
  We can describe the split $\C^n=\C^k\oplus \C^l$ that results from a pair of commuting complex structures explicilty. Indeed, because $J_1, J_2$ commute they can be simultaneously diagonalised. This means that all eigenvectors of $J_1$ of say eigenvalue $+\im$ are also eigenvectors of $J_2$. We then have
  \be
  E_+^1 = E^1_+ \cap E^2_- \oplus E^1_+ \cap E^2_+,
  \ee
  with the first factor consisting of eigenvectors of eigenvalue $+1$ of $J_1J_2$. 
  
\subsection{New characterisation of pure spinors of ${\rm Spin}(10)$}

We can now state and prove the following 
\begin{proposition} 
\label{pure-10}
A Weyl spinor of ${\rm Spin}(10)$ restricted to ${\rm Spin}(8)$ decomposes into a pair of Weyl spinors $\phi,\psi$ of ${\rm Spin}(8)$ of opposite parity. The ${\rm Spin}(10)$ spinor is then pure if and only if both ${\rm Spin}(8)$ spinors $\phi,\psi$ are pure and the complex structures $J_\phi, J_\psi$ on $\R^8$ they define commute. 
\end{proposition}
This characterisation of pure spinors of ${\rm Spin}(10)$ appears to be new. 

To prove this statement we use the notion of the product structure discussed above. Thus, as we discussed, given two pure spinors $\phi,\psi$ and their associated complex structures $J_\phi, J_\psi$ on $\R^8=\Oc$ we have two different decompositions of $\Oc_\C$:
\be
\Oc_\C = E_\phi \oplus \overline{E_\phi}, \qquad \Oc_\C = E_\psi \oplus \overline{E_\psi},
\ee
where we have denoted by $E_\phi, E_\psi$ the $+\im$ eigenspaces of $J_\phi, J_\psi$ respectively. Let us assume that the signs of the complex structures $J_\phi, J_\psi$ are chosen in such a way that $E_\phi, E_\psi$ coincide with the null subspaces $M(\phi),M(\psi)$ of $\phi,\psi$. These two complex structures commute if and only if each $E_\phi, E_\psi$ decomposes into the eigenspaces of the other complex structure, for example
\be\label{commuting-J}
E_\phi = E_1\oplus E_2, \qquad E_1 \subset E_\psi, \quad E_2\subset \overline{E_\psi}.
\ee
This means that we have the following decomposition into four null subspaces
\be\label{decomp-four}
\Oc_\C = E_1\oplus E_2 \oplus \overline{E_1}\oplus \overline{E_2},
\ee
with
\be
E_\phi = E_1\oplus E_2, \qquad E_\psi =  E_1 \oplus \overline{E_2}.
\ee
Thus, here $E_1$ is precisely the null subspace that is shared by both $M(\phi),M(\psi)$, and $E_2, \overline{E_2}$ are the complementary null subspaces on which $J_\phi, J_\psi$ disagree. The product structure 
$K=J_\phi J_\psi$ corresponding to the two commuting complex structures provides an orthogonal decomposition of $\Oc$ into two subspaces that are eigenspaces of $K$ of the opposite sign. In the case under consideration, the eigenspace of $K$ of one sign is the real subspace $E_1+\overline{E_1}$, and the eigenspace of the other sign is $E_2+\overline{E_2}$. It is clear that for a product structure that is constructed as a product of two commuting complex structures both eigenspaces of $K=J_\phi J_\psi$ are even dimensional. 

In general, for two distinct complex structures on $\R^8$, the possible dimensions of the space $E_1+\overline{E_1}$ can be $2, 4, 6$, which means that one, two and three null directions can be shared. The dimensions $0,8$ cannot occur for distinct complex structures. Indeed, when $E_\phi \cap E_\psi=0$, we have $E_\psi=\overline{E_\phi}$ and the two complex structures coincide. When $E_\phi=E_\psi$ the two complex structures again coincide. Further, the case when two null directions of $E_\phi, E_\psi$ are shared cannot arise in our setting of $J_\phi, J_\psi$ being defined by two ${\rm Spin}(8)$ pure spinors of opposite parity. Indeed, the only possible type of intersection allowed by Theorem \ref{thm:opposite} is in one and three null directions. 

We can now prove the proposition. In one direction, if we have two commuting complex structures $J_\phi, J_\psi$ on $\Oc$ of opposite parity, we have either ${\rm dim}_\C(E_\phi \cap E_\psi)=1$ or ${\rm dim}_\C(E_\phi \cap E_\psi)=3$. Let us choose the signs of the complex structures $J_\phi, J_\psi$ so that $E_\phi, E_\psi$ coincide with the null subspaces $E_\phi = M(\phi), E_\psi=M(\psi)$. Then when ${\rm dim}_\C(E_\phi \cap E_\psi)=3$ the null subspaces of $\phi, \psi$ intersect in three directions, and $B_1(\phi,\psi)=0$. This means that the compatibility conditions (\ref{beta2-beta1}) between the two pure spinors $\phi,\psi$ are satisfied and the pair $(\phi,\psi)$ makes up a pure spinor of ${\rm Spin}(10)$. 

When ${\rm dim}_\C(E_\phi \cap E_\psi)=1$ (and complex structures by assumption commute), we have
${\rm dim}_\C(E_\phi \cap \overline{E_\psi})=3$. This means that the null subspaces $M(\phi), M(R(\psi))$ intersect in three dimensions, which means that $B_1(\phi, R(\psi))=0$, which means that the pair of pure spinors $(\phi, R(\psi))$ makes up a pure spinor of ${\rm Spin}(10)$. 

To summarise, when the complex structures $J_\phi, J_\psi$ defined by two pure spinors of ${\rm Spin}(8)$ of opposite parity commute, either $(\phi,\psi)$ is a pure spinor of ${\rm Spin}(10)$, or $(\phi, R(\psi))$ is. 

In the opposite direction, let us assume that a pair $(\phi,\psi)$ of pure spinors of ${\rm Spin}(8)$ of opposite parity makes up a pure spinor of ${\rm Spin}(10)$. From (\ref{comput-XY}) we know that this implies $B_1(\phi,\psi)=0$, which in turn implies that ${\rm dim}_\C(M(\phi)\cap M(\psi)) = 3$. Further, as we have explicitly verified in (\ref{xi*}), the null subspaces of $\phi, R(\psi)$ intersect in one dimension. This means that we obtain the direct sum decomposition 
\be
M(\phi)=E_\phi = E_1\oplus E_2,
\ee 
where 
\be
E_1= M(\phi)\cap M(\psi), \qquad 
E_2 = M(\phi)\cap M(R(\psi))
\ee 
are complex dimension three and one respectively. 
We thus obtain the decomposition (\ref{decomp-four}), which means that the two complex structures with $+\im$ eigenspaces $E_\phi, E_\psi$ commute. Their product is a product structure giving the decomposition of $\Oc$ into the two dimensional subspace spanned by $\xi,\xi^*$ and the remaining six-dimensional subspace. 

As a corollary to this discussion we note that each of the pure spinors $\phi,\psi$ of ${\rm Spin}(8)$ has the stabiliser ${\rm SU}(4)$. Our discussion shows that the intersection of these stabilisers is ${\rm SU}(3)$, the one acting by unitary transformations of the three-dimensional space $E_1$ and preserving the top form on this space. Our characterisation of the SM gauge group as a subgroup of ${\rm Spin}(10)$, to be discussed below, is similar to this characterisation of ${\rm SU}(3)\subset{\rm Spin}(8)$. 

\subsection{Impure spinors in eight dimensions}

Equipped with understanding of the geometry of pure spinors, we can now understand the types of impure spinors that can arise in dimensions eight and ten.

Let us start in eight dimensions. We already saw that Majorana-Weyl spinors are impure because they are not null. It is also clear that complex spinors $\alpha_1+\im \alpha_2$ that are not null are impure. These are the only two possible types of impure spinors that can arise in this dimension, as we shall now explain.

The first elementary fact to note is that any non-null complexified octonion can be written as a sum of two null complexified octonions. To see this, it is convenient to first rescale the complexified octonion by a complex number to make it have norm one. It is clear that what we want to prove is invariant under such rescalings. After this is done, we have $\alpha_1+\im\alpha_2$ with $|\alpha_1|^2-|\alpha_2|^2=1$ and $(\alpha_1,\alpha_2)=0$. We can then define
\be
|\alpha_1|=\cosh\tau, \qquad |\alpha_2|= \sinh\tau,
\ee
so that 
\be\label{complex-impure}
\alpha_1+\im \alpha_2 =  \cosh\tau \frac{\alpha_1}{|\alpha_1|} + \im \sinh\tau \frac{\alpha_2}{|\alpha_2|}
= \\ \nonumber
\frac{1}{2}(\cosh\tau+\sinh\tau)  \left(\frac{\alpha_1}{|\alpha_1|} + \im \frac{\alpha_2}{|\alpha_2|}\right) +
\frac{1}{2}(\cosh\tau-\sinh\tau)  \left(\frac{\alpha_1}{|\alpha_1|} - \im \frac{\alpha_2}{|\alpha_2|}\right).
\ee
This means that any unit impure spinor of ${\rm Cl}(8)$ can be written as a linear combination of two complex conjugate pure spinors. For a complex impure spinors, the pure spinors arising in this decomposition are unique, as we have seen in (\ref{complex-impure}). The same applies for the case of a real octonion $\alpha_1$, except that this in this case the two pure spinors that can be used to represent $\alpha_1$ are not uniquely defined. Indeed, in this case we just need to select an (arbitrary) octonion $\alpha_2$ orthogonal to $\alpha_1$ and of the same norm. Then 
\be
\alpha_1= \frac{1}{2}(\alpha_1+\im \alpha_2) + \frac{1}{2}(\alpha_1-\im\alpha_2)
\ee
is a sum of two null complexified octonions. 

Given that we need only two pure spinors to represent an impure spinor in eight dimensions, we can understand the geometry of impure spinors. The MTN of a pure spinor in eight dimensions is four-dimensional. If we take two pure spinors $\phi,\psi$ of the same parity, their MTN's can intersect in four, two and zero dimensions. When they intersect in four dimensions $\phi$ is proportional to $\psi$. When they intersect in two dimensions, the sum is still a pure spinor. When they intersect in zero dimensions, the sum is impure. Thus, any impure spinor in eight dimensions is a sum of two pure spinors whose MTN's do not intersect. In terms of complexified octonions, the impure spinors are given by the linear combination of two null octonions that are complex conjugates of each other. 

The case of complex impure spinors is then different from the real case in terms of the geometry that arises. In the complex case the stabiliser of $\phi=\alpha_1+\im \alpha_2$ must stabilise both $\alpha_{1,2}$. It is then a copy of ${\rm Spin}(6)={\rm SU}(4)$ that stabilises the two complex conjugate pure spinors composing $\phi$. When the spinor is real $\phi=\alpha_1$, the stabiliser is different and is a copy of ${\rm Spin}(7)$ that stabilises the octonion $\alpha_1$. 

\subsection{Impure spinors in ten dimensions}

Let us now consider the more difficult case of impure spinors in ten dimensions. We first need the fact, proven in \cite{Charlton}, that an impure spinor in ten dimensions is a sum of two pure spinors, see Theorem 3.4.1 of this reference. Equipped with understanding that we do not need more than two pure spinors to decompose an impure spinor in ten dimensions, we can consider the possibilities that can arise. 

In ten dimensions, two distinct pure spinors of the same parity can intersect in three and one null directions. If they intersect in three their sum is still a pure spinor. So, an impure spinor $\phi$ in ten dimensions is a sum of two pure spinors that intersect in one dimension. We have already seen subsection \ref{sec:pure10} that an impure spinor $\phi$ defines a null vector $X+\im Y$. This null vector is the common null direction of the two pure spinor factors making up $\phi$. 

There are then just two types of impure spinors  because an impure spinor in dimension ten is an impure spinor in the eight dimensions that are orthogonal to $X,Y$. Its type is then the same as the type of the impure spinor in eight dimensions. And so the stabiliser of an impure spinor in ten dimensions is either ${\rm SU}(4)$ or ${\rm Spin}(7)$. 

To make this discussion more concrete, let us give examples of impure spinors of both types. Recall that a Weyl spinor in ten dimensions is a 2-component column with complexified octonions as entries. Then the spinor
\be
\left( \begin{array}{c} \textbf{1} \\ 0 \end{array}\right) 
\ee
is an example of an impure spinor whose stabiliser is ${\rm Spin}(7)$. On the other hand, the spinor
\be
\left( \begin{array}{c} \cosh\tau \textbf{1}+\im \sinh\tau {\bf u} \\ 0 \end{array}\right) 
\ee
is an example of a unit complex spinor whose stabiliser is ${\rm SU}(4)$. A different way to arrive at the types of spinor orbits in ten dimensions is explained in \cite{Bryant}. 

\section{GUT Models}
\label{sec:GUT}

We are now equipped with understanding that allows us to explain how the gauge group of the Standard model
\be
G_{\rm SM} = {\rm SU}(3)\times{\rm SU}(2)\times {\rm U}(1)_Y
\ee
sits inside the gauge group of Grand Unified Theory (GUT) ${\rm Spin}(10)$. Here $Y$ stands for the hypercharge. It is known, see e.g. \cite{Baez:2009dj}, that $G_{\rm SM}$ can be described as the intersection of the Georgi-Glashow ${\rm SU}(5)\subset {\rm Spin}(10)$ and the Pati-Salam ${\rm SO}(4)\times{\rm SO}(6)\subset {\rm SO}(10)$ subgroups, when these are appropriately aligned. The purpose of this section is to explain this statement, and also explain the geometric meaning of the "appropriately aligned". 

Also, once we specify how $G_{\rm SM}$ fits inside ${\rm Spin}(10)$, we can identify various vectors in the Weyl representation $\C^{16}$ of ${\rm Spin}(10)$ with elementary particles. Given that in our octonionic model of ${\rm Spin}(10)$ the Weyl representation gets identified with $\Oc_\C^2$, choosing an embedding of $G_{\rm SM}$ inside ${\rm Spin}(10)$ identifies elementary particles with various complexified octonions. The arising dictionary is explained in this section. 

\subsection{Particles}

To explain how $G_{\rm SM}$ sits inside ${\rm SU}(5)$ and why $G_{\rm SM}$ is related to the intersection of ${\rm SU}(5)$ with the Pati-Salam group ${\rm SO}(4)\times{\rm SO}(6)$, we need to start with the recap of the Standard Model (SM) elementary particles and their properties. 

Particles of the SM come in three generations that are exact copies of each other as far as the $G_{\rm SM}$ transformation properties of the particles are concerned. We will only consider a single generation in this paper. 

The particles of one generation can be described as sixteen 2-component Lorentz group ${\rm Spin}(3,1)$ spinors. This group has two different type of spinor representations (two different types of Weyl spinors). Spinors of one type transform under ${\rm Spin}(3,1)={\rm SL}(2,\C)$ as complex conjugates of spinors of the other type. The Dirac Lagrangian is real, and involves one 2-component spinor for each particle, as well as the complex conjugate spinor. It is then convenient to describe all particles using the 2-component Lorentz spinors of the same type, keeping it in mind that the complex conjugate spinors of the opposite type also appear in the Lagrangian. 

Particles are distinguished according to their transformation properties with respect to $G_{\rm SM}$. We have the following table, with the convention that we use 2-component Lorentz spinors of the same type to describe all particles, see  \cite{Dreiner:2008tw} for more details
\begin{center}\label{table}
\begin{tabular}{ |c|c|c| c|c|c|} 
 \hline
 Particles & SU(3) & SU(2)  & $Y$ & $T^3$ & $Q=T^3+Y$ \\ 
 \hline
 $Q=\left( \begin{array}{c} u \\ d\end{array}\right)$ & triplet & doublet & 1/6  & $\begin{array}{c} 1/2 \\ -1/2\end{array}$ & $\begin{array}{c} 2/3 \\ -1/3\end{array}$ \\
 $\begin{array}{c} \bar{u} \\ \bar{d}\end{array}$ & anti-triplet & singlet & $\begin{array}{c} -2/3 \\ 1/3\end{array}$  & 0 & $\begin{array}{c} -2/3 \\ 1/3\end{array}$ \\
 $L=\left( \begin{array}{c} \nu \\ e \end{array}\right)$ & singlet & doublet & -1/2  & $\begin{array}{c} 1/2 \\ -1/2\end{array}$ & $\begin{array}{c} 0 \\ -1\end{array}$ \\
$ \bar{e}$ & singlet & singlet & 1 & 0 & 1 \\
 \hline
\end{tabular}
\end{center}
\bigskip

The particles that are non-singlets with respect to the strong force ${\rm SU}(3)$, namely $Q, \bar{u}, \bar{d}$ are called quarks. The remaining particles $L, \bar{e}$ are called leptons. Two different 2-component spinors are needed to describe a particle with the same name (for all particles but the neutrino), e.g. both $u,\bar{u}$ are needed to describe the up quark {\bf and} its anti-particle. One can say that $u$ describes the up quark while $\bar{u}$ describes its anti-particle, although this terminology is empty unless one defines an independent notion of particles and anti-particles. We do not need this and will not attempt such a definition in this paper. The other pairs are $d,\bar{d}$ that describe the down quark and its anti-particle, and $e,\bar{e}$ that describe the electron and its anti-particle (positron). The only 2-component spinor that does not have its partner is $\nu$, and this is because in the SM neutrino coincides with its anti-particle. However, the embedding into ${\rm Spin}(10)$ predicts $\bar{\nu}$ as well. 

\subsection{Georgi-Glashow ${\rm SU}(5)$ GUT}

The Georgi-Glashow ${\rm SU}(5)$ GUT arises from the realisation that one can embed the rank 4 $G_{\rm SM}$ into the rank 4 simple Lie group ${\rm SU}(5)$. The embedding is as follows
\be\label{23-embedding}
 {\rm SU}(3)\times{\rm SU}(2)\times{\rm U}(1)_Y \ni (h_3, h_2, \alpha) \to \left( \begin{array}{cc} \alpha^2 h_3 & 0 \\ 0 & \alpha^{-3} h_2 \end{array}\right)\in {\rm SU}(5).
 \ee
 We can then attempt to identify the $\C^5$ representation of ${\rm SU}(5)$ with particles. It is clear that as the representation of ${\rm SU}(3)\times{\rm SU}(2)$ $\C^5$ will split into $\C^3\oplus\C^2$, i.e. a triplet and a doublet. It is also clear that the one will have $3Y_{\C^3} + 2Y_{\C^2}=0$. The only particle doublet that could be identified with $\C^2$ is $L$, which has the $Y$ charge of $-1/2$. The other particle then must have the $Y$ charge of $1/3$, and thus be identified with $\bar{d}$. Because $\bar{d}$ transforms as an anti-triplet, we must correct the identification and identify $L,\bar{d}$ as components of $\bar{C}^5$ instead. All in all, the requirement that $\bar{\C}^5$ of ${\rm SU}(5)$ can be identified with particles under the embedding of the type (\ref{23-embedding}) fixes the embedding to be
 \be\label{23-embedding*}
 {\rm SU}(3)\times{\rm SU}(2)\times{\rm U}(1)_Y \ni (h_3, h_2, e^{\im\phi}) \to \left( \begin{array}{cc} e^{-\im\phi/3} h_3 & 0 \\ 0 & e^{\im\phi/2} h_2 \end{array}\right)\in {\rm SU}(5),
 \ee
 with the convention that the 5-component columns that such matrices act on are in $\C^5$, and the complex conjugate matrices act on $\bar{\C}^5$. We then get the desired $\bar{\C}^5=\bar{\C}^3\oplus\bar{\C}^2$ and the correct pattern of the $Y$-charges. 
 
 The other particles are identified as follows. We consider the representation $\Lambda^3\bar{\C}^5$. It splits as
 \be
  \Lambda^3(\bar{\C}^3 \oplus \bar{\C}^2)=  \Lambda^3 \bar{\C}^3 \oplus \Lambda^2 \bar{\C}^3 \otimes \bar{\C}^2 \oplus \bar{\C}^3\otimes \Lambda^2 \bar{\C}^2. 
  \ee
  The first factor here has the $Y$-charge of $+1$ and is an ${\rm SU}(3)$ singlet. It must be identified with the partner $\bar{e}$ of the electron. The second factor is an ${\rm SU}(2)$ doublet of $Y$-charge $2/3-1/2=1/6$, which is $Q$. The third factor is an ${\rm SU}(3)$ triplet with $Y$-charge $1/3-1=-2/3$, which is $\bar{u}$. This finishes the assignment of particles to representations of ${\rm SU}(5)$. 
  
Having considered $\Lambda^1\bar{\C}^5, \Lambda^3\bar{\C}^5$, it is very natural to consider also the representation $\Lambda^5\bar{\C}^5$. It splits as
 \be
 \Lambda^5\bar{\C}^5 = \Lambda^3\bar{\C}^3 \otimes \Lambda^2 \bar{\C}^2.
 \ee
 Both factors are ${\rm SU}(3)$ and ${\rm SU}(2)$ singlets, and the total $Y$ charge is zero. This representation can be used to describe the partner $\bar{\nu}$ of the neutrino that is missing from the above table but must be considered in the ${\rm Spin}(10)$ GUT. 
 
 The preceding discussion can be summarised as follows. The particles of the Table \ref{table} together with $\bar{\nu}$ become identified with the following representations of ${\rm SU}(5)$
 \be\label{split-GG}
 \Lambda^1 \bar{C}^5 \oplus \Lambda^3 \bar{\C}^5 \oplus \Lambda^5 \bar{\C}^5= (\bar{C}^3\oplus \bar{C}^2) \oplus (\Lambda^3 \bar{\C}^3 \oplus \Lambda^2 \bar{\C}^3 \otimes \bar{\C}^2 \oplus \bar{\C}^3\otimes \Lambda^2 \bar{\C}^2) \oplus (\Lambda^3\bar{\C}^3 \otimes \Lambda^2 \bar{\C}^2)= \\ \nonumber
 (\bar{d}, L) + (\bar{e}, Q, \bar{u}) + (\bar{\nu}).
 \ee
 The vector space on the left-hand side is the space of odd degree differential forms on $\bar{C}^5$. As we know from our description of Section \ref{sec:2n}, this space is also one of the two Weyl spinor representations of ${\rm Spin}(10)$, of which ${\rm SU}(5)$ is naturally a subgroup. So, while the particles of the SM are only partially unified by several different representations of ${\rm SU}(5)$, they become fully unified into a single irreducible representation of ${\rm Spin}(10)$ GUT. 
 
In our discussion above we have associated particles with various vectors in the space of odd degree polyforms on $\bar{C}^5$. For the discussion that follows, it will be more convenient to describe particles using the space of even degree differential forms on $\C^5$ instead. This is the space
 \be
 \Lambda^4 \C^5+ \Lambda^2\C^5 + \Lambda^0 \C^5 
 \ee
 The first factor here decomposes as
 \be
 \Lambda^4(\C^3\oplus\C^2) = \Lambda^3 \C^3 \otimes \C^2 \oplus \Lambda^2 \C^3 \otimes \Lambda^2 \C^2.
 \ee
 The first term here is an ${\rm SU}(3)$ singlet and ${\rm SU}(2)$ doublet of $Y$ charge $-1+1/2=-1/2$, which is $L$. The second term is an ${\rm SU}(3)$ triplet and ${\rm SU}(2)$ singlet of $Y$ charge $-2/3+1= 1/3$, which must be identified with $\bar{d}$. 
 
 The factor $\Lambda^2 \C^5$ decomposes as
 \be\label{5-23-split}
 \Lambda^2 \C^5 = \Lambda^2( \C^3+\C^2) = \Lambda^2 \C^3\oplus \C^3\otimes \C^2 \oplus\Lambda^2 \C^2. 
 \ee
 The representation $\Lambda^2 \C^2$ is the ${\rm SU}(2)$ singlet with the $Y$-charge 1, and so needs to be identified with $\bar{e}$. The representation $\C^3\otimes \C^2$ has the $Y$-charge $ -1/3+ 1/2=1/6$, and is a triplet and a doublet. So, it needs to be identified with $Q$. The last remaining factor $\Lambda^2 \C^3$ is a triplet and has the $Y$-charge of $-2/3$, and is identified with $\bar{u}$. 
 
 This gives the following alternative assignment of particles with components of even degree polyforms
 \be\label{part-assign-even}
 \Lambda^4 \C^5+ \Lambda^2\C^5 + \Lambda^0 \C^5 = (\Lambda^3 \C^3 \otimes \C^2 \oplus \Lambda^2 \C^3 \otimes \Lambda^2 \C^2) \oplus (\Lambda^2 \C^3\oplus \C^3\otimes \C^2 \oplus\Lambda^2 \C^2) \oplus (\Lambda^0\C^3 \otimes\Lambda^0 \C^2) =\\ \nonumber
 (L,\bar{d}) + ( \bar{u}, Q, \bar{e}) + (\bar{\nu}). 
 \ee
 Alternatively, we can write this as 
 \be \nonumber
 \Lambda^0\C^3 \otimes\Lambda^0 \C^2\leftrightarrow \bar{\nu}, \\ \nonumber
 \Lambda^2 \C^3 \otimes\Lambda^0 \C^2 \leftrightarrow \bar{u}, \\ 
\label{part-assign*}  \Lambda^0\C^3 \otimes \Lambda^2\C^2 \leftrightarrow \bar{e}, \\ \nonumber
 \Lambda^2 \C^3 \otimes \Lambda^2\C^2 \leftrightarrow \bar{d}, \\ \nonumber
  \Lambda^1\C^3 \otimes \Lambda^1\C^2 \leftrightarrow Q, \\ \nonumber
 \Lambda^3 \C^3 \otimes \Lambda^1\C^2\leftrightarrow L.
 \ee

\subsection{Pati-Salam ${\rm SO}(4)\times{\rm SO}(6)$ GUT}

Another perspective on the unification is provided by the Pati-Salam ${\rm SO}(4)\times{\rm SO}(6)$. In fact, it is better to talk about the group ${\rm Spin}(4)\times{\rm Spin}(6)={\rm SU}(2)\times{\rm SU}(2)\times{\rm SU}(4)$. 

The SM particles are assigned to representations of ${\rm SU}(2)\times{\rm SU}(2)\times{\rm SU}(4)$ following the idea that leptons and quarks are part of a bigger structure with the group ${\rm SU}(4)$ acting on all these particles and mixing them together. Leptons are then just the fourth colour of quarks. One also puts all the barred 2-component spinors in doublets of the second ${\rm SU}(2)$. To do this for $\bar{e}$ necessitates introducing $\bar{\nu}$. Thus, unlike ${\rm SU}(5)$, the Pati-Salam GUT requires the neutrino antiparticle $\bar{\nu}$ and thus predicts it. 

One gets the following assignment of particles to representations of ${\rm SU}(2)\times{\rm SU}(2)\times{\rm SU}(4)$
\be\label{split-PS}
(\C^2,0,\C^4) \oplus (0,\C^2,\bar{\C}^4) = (Q,L) + (\bar{Q},\bar{L}),
\ee
where $0$ stands for the trivial representation, it is understood that $\bar{Q}$ is a 2-component ${\rm SU}(2)$ spinor with components $\bar{d},\bar{u}$, and similarly $\bar{L}$ is a 2-component spinor with components $\bar{e},\bar{\nu}$. It is also natural to put some additional signs in this identification, see e.g. \cite{DiLuzio:2011mda}, Section 1.3.1, but this will not be important for us, as everything will be fixed by the eventual embedding into ${\rm Spin}(10)$, to be described below. 

We note that (\ref{split-PS}) is precisely the branching rule for any of the two Weyl representations of ${\rm Spin}(10)$ under the subgroup  ${\rm Spin}(4)\times{\rm Spin}(6)\subset{\rm Spin}(10)$. This statement needs clarification, as it depends on the embedding. The embedding for which the branching rule (\ref{split-PS}) holds is the one for which ${\rm SO}(4)\times{\rm SO}(6)\subset {\rm SO}(10)$ are the two block-diagonal $4\times 4$ and $6\times 6$ orthogonal blocks of the $10\times 10$ orthogonal matrix. 

The fact that the branching rule is as stated is seen very easily using the creation/annihilation model of ${\rm Cl}(10)$. Indeed, starting by selecting a complex structure on $\R^{10}$ to identify $\R^{10}=\C^5$, we get 5 pairs of creation/annihilation operators that generate ${\rm Cl}(10)$ according to (\ref{Gamma-a}). If we choose a split
\be
\C^5=\C^2\oplus \C^3,
\ee
we get two pairs of creation/annihilation operators generating ${\rm Cl}(4)$, and another three pairs generating ${\rm Cl}(6)$. The corresponding ${\rm Spin}(4),{\rm Spin}(6)$ groups have their images in ${\rm SO}(10)$ precisely the diagonal $4\times 4$ and $6\times 6$ orthogonal blocks. It is then easy to see that any of the two Weyl representations of ${\rm Spin}(10)$ splits as in (\ref{split-PS}). Indeed, the space of e.g. even polyforms on $\C^5$ splits as 
\be
\Lambda^{\rm even} \C^5= \Lambda^{\rm even} \C^2 \otimes \Lambda^{\rm even} \C^3 \oplus \Lambda^{\rm odd} \C^2 \otimes \Lambda^{\rm odd} \C^3.
\ee
Taking into account that $\Lambda^{\rm even/odd} \C^3$ coincide with the representations $\C,\bar{\C}$ of ${\rm Spin}(6)={\rm SU}(4)$, and the fact that $\Lambda^{\rm even/odd} \C^2$ are the two doublet representations of the two different ${\rm SU}(2)$ in ${\rm Spin}(4)={\rm SU}(2)\times{\rm SU}(2)$, we recover the branching rule (\ref{split-PS}).

\subsection{$G_{\rm SM}$ as the intersection of the Georgi-Glashow and Pati-Salam groups}

Having understood that the SM gauge group $G_{\rm SM}$ can be thought as a subgroup of both the Georgi-Glashow group ${\rm SU}(5)$ as well as the Pati-Salam group ${\rm Spin}(4)\times{\rm Spin}(6)$, the natural question to ask is whether $G_{\rm SM}\subset{\rm Spin}(10)$ can be characterised in terms of the intersection of these two subgroups. The answer is in the affirmative. To describe the corresponding statement, let us start by introducing some useful terminology.

We start by choosing a complex structure $J$ on $\R^{10}$, thus identifying $\R^{10}=\C^5$, together with a holomorphic top form $\Omega$ on $\Lambda\C^5$. This chooses a ${\rm SU}(5)\subset {\rm Spin}(10)$. As discussed previously, the data $J,\Omega$ that went into this choice is in one-to-one correspondence with a pure spinor $\Psi$ of ${\rm Spin}(10)$
\be
\Psi \sim (J,\Omega).
\ee
It is useful to indicate this explicitly and write ${\rm SU}(5)_\Psi\equiv {\rm SU}(5)_{J,\Omega}$ for the ${\rm SU}(5)$ that corresponds to the pure spinor $\Psi$. Note that ${\rm SU}(5)_\Psi$ is just the stabiliser subgroup of $\Psi$ in ${\rm Spin}(10)$.
\begin{definition} We call a Pati-Salam subgroup ${\rm Spin}(4)\times{\rm Spin}(6)\subset{\rm Spin}(10)$ {\bf J-compatible} if it arises from a split $\C^5=\C^2\oplus\C^3$, by splitting the creation/annihilation operators generating ${\rm Cl}(10)$ into two groups, one consisting of two pairs and generating ${\rm Cl}(4)$, and the other consisting of three pairs and generating ${\rm Cl}(6)$, as described in the previous subsection. 
\end{definition}

\begin{proposition} The SM gauge group $G_{\rm SM}$ is precisely the intersection of a Georgi-Glashow ${\rm SU}(5)_{J,\Omega}$ with a J-compatible Pati-Salam subgroup ${\rm Spin}(4)\times{\rm Spin}(6)\subset{\rm Spin}(10)$.
\end{proposition}
\begin{remark} This fact was noticed and described in \cite{Baez:2009dj}. But there are likely earlier references containing it.
\end{remark}

A proof of this proposition is implicit in the description of the embedding of $G_{\rm SM}$ into ${\rm SU}(5)$ specified by the equation (\ref{23-embedding}). Indeed, matrices appearing in the image of this embedding are precisely those ${\rm SU}(5)$ matrices that preserve the split $\C^5=\C^2\oplus\C^3$ of the fundamental representation $\C^5$. This makes it clear that $G_{\rm SM}$ is precisely the subgroup of ${\rm SU}(5)$ that preserves this split. 

On the other hand, a general, not necessarily J-compatible Pati-Salam subgroup of ${\rm Spin}(10)$ arises by selecting a split $\R^{10}=\R^4\oplus \R^6$. The geometric structure that encodes such a split was already described above, and is that of a product structure. Indeed, a choice of ${\rm Spin}(4)\times{\rm Spin}(6)\subset{\rm Spin}(10)$ is equivalent to a choice of the split $\R^{10}=\R^4\oplus \R^6$. Such a choice is encoded by a product structure $K$ that has eigenvalue of one sign on $\R^4$ and eigenvalue of another sign on $\R^6$. The corresponding Pati-Salam subgroup is only J-compatible if the split $\R^{10}=\R^4\oplus \R^6$ is compatible with the selected complex structure $J$ on $\R^{10}$, i.e. comes from a split $\C^5=\C^2\oplus\C^3$.  

\section{Particle dictionary}
\label{sec:particles}

Having understood how the SM gauge group sits inside ${\rm Spin}(10)$, we can work out the dictionary identifying various components of a Weyl spinor of ${\rm Spin}(10)$ with elementary particles.

Recall that we have given two descriptions of the spinors of ${\rm Spin}(10)$. In the description based on creation/annihilation operators Weyl spinors are even or odd polyforms on $\C^5$. In the model based on octonions, a Weyl spinor is a 2-component column with entries in complexified octonions. We will explain the particle identification in both of these models. 

Particles can only be mapped to components of a Weyl spinor of ${\rm Spin}(10)$ when an embedding of the SM gauge group is selected. As we discussed in the previous subsection, to fix an embedding we need to select both a Georgi-Glashow ${\rm SU}(5)$, which amounts to choosing a pure spinor of ${\rm Spin}(10)$, as well as a J-compatible Pati-Salam ${\rm Spin}(4)\times{\rm Spin}(6)$, or, equivalently, a split $\C^5=\C^2\oplus\C^3$. 

\subsection{Choice of Georgi-Glashow}
\label{sec:GG}

Let us start by choosing a pure spinor of ${\rm Spin}(10)$. As described in Proposition \ref{pure-10}, a pure spinor of ${\rm Spin}(10)$ can be described as two pure spinors of ${\rm Spin}(8)$ of opposite parity, that are moreover suitably compatible. The simplest possibility is to take one of these two pure spinors to be equal to zero. In this case the compatibility conditions are automatically satisfied. A positive parity Weyl spinor of ${\rm Spin}(10)$ is a 2-component column 
\be
\Psi_1=\left( \begin{array}{c} \alpha \\ \beta\end{array}\right), \qquad \alpha,\beta\in\Oc_\C.
\ee
Recall that the complexified octonion $\alpha$ encodes a positive parity polyform on $\C^4$, and $\beta$ encodes a negative parity polyform, see (\ref{psi-pm-10}). We will take the pure spinor fixing ${\rm SU}(5)$ to be one with $\beta=0$. Then $\alpha$ is just a pure spinor of ${\rm Spin}(8)$, which is a null complexified octonion. Given that the imaginary octonionic unit ${\bf e}^4={\bf u}$ plays a special role in our description (as being identified with the identity quaternion in ${\rm Im}(\Oc)=\Hq\oplus{\rm Im}(\Hq)$), we take $\alpha={\bf 1} +\im {\bf u}$, so that the pure spinor of ${\rm Spin}(10)$ whose stabiliser is ${\rm SU}(5)$ is given by
\be\label{psi-special}
\Psi_1=\left( \begin{array}{c} {\bf 1} +\im {\bf u}  \\ 0 \end{array}\right), \qquad \alpha,\beta\in\Oc_\C.
\ee
This fixes the Georgi-Glashow subgroup of ${\rm Spin}(10)$, and any other choice is conjugate to this one inside ${\rm Spin}(10)$. We will see that this is the most natural choice of the pure spinor and the associated complex structure on $\R^{10}$, given all other conventions adopted. 

It will be useful to be explicit, and describe both the complex structure on $\R^{10}$ that this pure spinor corresponds to, as well as the top holomorphic form. Both are recovered by considering various differential forms on $\R^{10}$ that are quadratic in $\psi$ (and possibly its complex conjugate). Let us first recover the complex structure. This is done by considering the 2-form on $\R^{10}$ arising as $\langle R(\Psi_1), \Gamma\Gamma \Psi_1\rangle$. Using the explicit form of the $\Gamma$-matrices (\ref{gamma-x}) and (\ref{gamma-910}), we get the following components
\be
\langle R(\Psi_1), \Gamma^x \Gamma^y \Psi_1\rangle  = ((\alpha_1 -\im \alpha_2), L_{\bar{x}} L_y (\alpha_1+\im \alpha_2)) + ((\beta_1 -\im \beta_2), L_x L_{\bar{y}} (\beta_1+\im \beta_2)), 
\\ \nonumber
\langle R(\Psi_1), \Gamma^x \Gamma^{9} \Psi_1\rangle=- ( (\alpha_1 -\im \alpha_2), L_{\bar{x}} (\beta_1+\im \beta_2)) + ((\beta_1 -\im \beta_2), L_x (\alpha_1+\im \alpha_2)), 
\\ \nonumber
\langle R(\Psi_1), \Gamma^x \Gamma^{10} \Psi_1\rangle = \im ( (\alpha_1 -\im \alpha_2), L_{\bar{x}} (\beta_1+\im \beta_2)) + \im ((\beta_1 -\im \beta_2), L_x (\alpha_1+\im \alpha_2), 
  \\ \nonumber
\langle R(\Psi_1), \Gamma^9 \Gamma^{10} \Psi_1\rangle= \im ( (\alpha_1 -\im \alpha_2),  (\alpha_1+\im \alpha_2)) -\im ((\beta_1 -\im \beta_2), (\beta_1+\im \beta_2)).
\ee
When $\alpha={\bf 1}+\im{\bf u}, \beta=0$ the non-zero components become
\be
\langle R(\Psi_1), \Gamma^x \Gamma^y \Psi_1\rangle =   (({\bf 1}-\im{\bf u}),  L_{\bar{x}} L_y ({\bf 1}+\im{\bf u}))
\\ \nonumber
\langle R(\Psi_1), \Gamma^9 \Gamma^{10} \Psi_1\rangle= 2 \im.
\ee
This gives
\be
\langle R(\Psi_1), \Gamma \Gamma \Psi_1\rangle = 2\im (  e^{15} +e^{26} + e^{37} + e^{48} + e^{9\,10}).
\ee
This is a multiple of the Kahler form for the complex structure whose eigendirections are
\be\label{eigendir}
e^4+\im e^8, \quad e^1+\im e^5, \quad e^2+\im e^6, \quad e^3+\im e^7, \quad e^9+\im e^{10}.
\ee
We remind the reader that the polyform construction of the spinor representations is based on the identification of these null directions in $\R^{10}_\C$ with the 1-forms $dz_I, I=1,\ldots, 5$, see Section \ref{sec:2n}.

Another computation gives the complex 5-form $B_5(\Psi_1)$. We have
\be
\langle \Psi_1, \Gamma\Gamma\Gamma\Gamma\Gamma \Psi_1\rangle=
2 ( e^9 +\im e^{10}) \wedge (e^4+\im e^8)\wedge (e^1+\im e^5)\wedge (e^2+\im e^6) \wedge (e^3+\im e^7),
\ee
which is of course a multiple of the wedge product of the directions in (\ref{eigendir}), which are the null directions of the pure spinor (\ref{psi-special}). 

\subsection{Choice of a compatible Pati-Salam}

The second step is to choose a J-compatible Pati-Salam subgroup of ${\rm Spin}(10)$, or equivalently a split $\C^5=\C^2\oplus\C^3$. In the octonionic description of ${\rm Spin}(10)$, once an imaginary unit ${\bf u}$ is chosen to get the pure spinor ${\bf 1}+\im{\bf u}$ of ${\rm Spin}(8)$, the most natural $\C^5=\C^2\oplus\C^3$ split is given by
\be\label{c2-c3}
\C^2\oplus\C^3 = {\rm Span}(e^4+\im e^8, e^9+\im e^{10}) \oplus {\rm Span}(e^1+\im e^5, e^2+\im e^6, e^3+\im e^7)=\\ \nonumber
{\rm Span}(dz_4, dz_5) \oplus{\rm Span}(dz_1,dz_2,dz_3).
\ee
So, we make the corresponding choice of the J-compatible Pati-Salam subgroup of ${\rm Spin}(10)$. This fixes both the Georgi-Glashow and Pati-Salam subgroups, and thus the SM gauge group inside ${\rm Spin}(10)$ as their intersection. In principle, this gives everything one needs to spell out the identification with the elementary particles. However, it will be useful to develop few other tools that make the identification with particles more transparent. 

\subsection{Complex structure on spinors}

We now note that a choice of a Pati-Salam ${\rm Spin}(4)\times{\rm Spin}(6)$ subgroup of ${\rm Spin}(10)$ gives rise to a complex structure on the space of Weyl spinors. Indeed, it is clear that a choice of ${\rm Spin}(4)\times{\rm Spin}(6)\subset{\rm Spin}(10)$ is equivalent to a choice of the split $\R^{10}=\R^{4}\oplus\R^6$. Once a choice like this is made, we can consider the product of six $\Gamma$-matrices in the directions $\R^6$. With the choice (\ref{c2-c3}) this is the operator
\be\label{product-six-alt}
J_S := \Gamma_1 \Gamma_2\Gamma_3 \Gamma_5 \Gamma_6 \Gamma_7.
\ee
It is easy to see that $J_S^2=-\id$. Moreover, being a product of an even number of $\Gamma$-matrices, this is an operator that preserves the spaces of Weyl spinors. So, this is a complex structure on both of the spaces $S_\pm$. A calculation gives
\be\label{J-Lu}
J_S\Big|_{S_+} = -\left( \begin{array}{cc} L_{\bf u} & 0 \\ 0 & L_{\bf u} \end{array}\right).
\ee
It is understood here that spinors in $S_+$ are 2-component columns with entries in $\Oc_\C$, and the complex structure $J_S$ acts on each of the $\Oc_\C$ entries as $L_{\bf u}$. This complex structure played the crucial role in some related earlier developments, see \cite{Krasnov:2019auj}, \cite{Krasnov:2021jjp}.

A choice of the split $\R^{10}=\R^{4}\oplus\R^6$ produces a complex structure $J_S$, but it is also easy to see that the complex structure $J_S$ completely encodes the information about the split. Indeed, it is easy to check that the ${\rm Spin}(10)$ transformations that commute with $J_S$ are precisely those in ${\rm Spin}(4)\times{\rm Spin}(6)$. Indeed, it is most transparent to see this at the level of the Lie algebra. The Lie algebra $\mathfrak{spin}(10)$ is generated by products of pairs of distinct $\Gamma$-matrices. The only Lie algebra elements that commute with the product (\ref{product-six-alt}) of six Clifford generators are either products of $\Gamma$ matrices that appear in (\ref{product-six-alt}), or products of those that do not appear. This shows that the commutant of $J_S$ in $\mathfrak{spin}(10)$ is precisely $\mathfrak{spin}(4)\times \mathfrak{spin}(6)$. Thus, a complex structure $J_S$ on $S_\pm$ (of a certain type that we will characterise later) encodes the information about a Pati-Salam ${\rm Spin}(4)\times{\rm Spin}(6)$ subgroup of ${\rm Spin}(10)$. 

It will be useful for what follows to describe the eigenspaces of $J_S$ on $S_+$ explicitly. The space of Weyl spinors $S_+$ is a copy of $\C^{16}$. The eigenspaces of $J_S$ are half-dimensional, and so copies of $\C^8$. The complex structure $L_{\bf u}$ acting on $\Oc_\C$ splits the space of complexified octonions into the eigenspaces of $L_{\bf u}$ of eigenvalue $\pm\im$. Let us denote these eigenspaces by $E_\pm$, with the convention that $E_-$ is the eigenspace of eigenvalue $-\im$. Each $E_\pm=\C^4$ and for the complex structure (\ref{J-Lu}) we have
\be
E_+= {\rm Span}( {\bf 1}+\im {\bf u}, {\bf e}^1+\im {\bf e}^5, {\bf e}^2+\im {\bf e}^6, {\bf e}^3+\im {\bf e}^7),
 \\ \nonumber
E_-={\rm Span}( {\bf 1}-\im {\bf u}, {\bf e}^1-\im {\bf e}^5, {\bf e}^2-\im {\bf e}^6, {\bf e}^3-\im {\bf e}^7).
\ee

Then 
\be
\Oc_\C = E_+ \oplus E_-.
\ee
It is then clear that any spinor in $S_+$ splits under $J_S$ as
\be\label{split-Sp}
\left( \begin{array}{c} \alpha \\ \beta \end{array}\right) = \left( \begin{array}{c} \alpha_+ \\ \beta_+ \end{array}\right) +  \left( \begin{array}{c} \alpha_- \\ \beta_- \end{array}\right), \qquad \alpha_+,\beta_+\in E_+, \quad \alpha_-,\beta_-\in E_-.
\ee
The transformations from ${\rm Spin}(4)\times{\rm Spin}(6)$ commute with $J_S$ and thus preserve this splitting of $S_+$. 

It is useful to re-interpret the split in (\ref{split-Sp}) in terms of polyforms. We recall from our discussion in (\ref{octonions-polyforms-even}) that complexified octonions $\alpha_+$ are precisely the even polyforms in $\Lambda(\C^3)$, and the octonions in $\alpha_-$ are the odd polyforms in $\Lambda(\C^3)$ wedged with $dz_4$. Similarly, from (\ref{octonions-polyforms-odd}) we recall that the octonions $\beta_+$ are the even polyforms in $\Lambda(\C^3)$ wedged with $dz_4$, and octonions $\beta_-$ are odd polyforms in $\Lambda(\C^3)$. And so, we can rewrite the split (\ref{split-Sp})  in terms of polyforms
\be\label{sp-polyforms}
\left( \begin{array}{c} \alpha \\ \beta \end{array}\right)  \leftrightarrow \alpha_+ + \beta_+ \wedge dz_{45} + \alpha_-\wedge dz_4 + \beta_-\wedge dz_{5}, \qquad \alpha_+,\beta_+\in \Lambda^{even}(\C^3), \quad 
\alpha_-,\beta_-\in \Lambda^{odd}(\C^3).
\ee
However, we can recall from our discussion of the spinors of ${\rm Spin}(4)$ that the spinor $\alpha_+ + \beta_+ \wedge dz_{45}$ is an even spinor, and $\alpha_-\wedge dz_4 + \beta_-\wedge dz_{5}$ is an odd spinor of this group. This shows that the split (\ref{split-Sp}), in terms of representations of ${\rm Spin}(4)\times{\rm Spin}(6)={\rm SU}(2)\times{\rm SU}(2)\times{\rm SU}(4)$ is the following split
\be
S_+ = (\Lambda^{even}(\C^3)\otimes \Lambda^{even}(\C^2)) \oplus (\Lambda^{odd}(\C^3)\otimes \Lambda^{odd}(\C^2)).
\ee
This matches (\ref{split-PS}). 

\subsection{Dictionary}

All in all, we get the desired dictionary translating spinors of ${\rm Spin}(10)$ as 2-component columns with complexified octonion entries into particles. We see that the central element of this dictionary is the complex structure $(\ref{J-Lu})$ on spinors, which splits the space of Weyl spinors into two orthogonal subspaces (\ref{split-PS}), and encodes the choice of the Pati-Salam subgroup. 

Restriction to the Georgi-Glashow subgroup introduces further structure, and allows to complete identification with particles. The Georgi-Glashow subgroup acts as the group of unitary transformations mixing the basic 1-forms $dz_I, I=1,\ldots 5$. Its intersection with Pati-Salam introduces the split (\ref{c2-c3}), and thus allows only the transformations mixing $dz_{1,2,3}$ and $dz_{4,5}$. The identification with particles is then given by (\ref{part-assign-even}). We spell it out. 

 The Georgi-Glashow group ${\rm SU}(5)$ does not mix polyforms of different degrees, and so the spaces $\Lambda^{even,odd}\C^3$ split as
 \be
 \Lambda^{even}(\C^3) = \Lambda^0 \C^3\oplus \Lambda^2 \C^3, \qquad
  \Lambda^{odd}(\C^3) =\Lambda^1 \C^3\oplus \Lambda^3 \C^3,
  \ee
 Also, only $\Lambda^{odd}(\C^2)$ transforms as a doublet under the weak ${\rm SU}(2)\subset{\rm SU}(5)$. The space  $\Lambda^{even}(\C^2)=\Lambda^0(\C^2)\oplus\Lambda^2(\C^2)$ transforms as two singlets. The complete identification of particles with components of $S_+$ as polyforms is then given by (\ref{part-assign*}).
  
 We can now compute the mapping of particles to complexified octonions. The polyforms $1\in \Lambda^0\C^3$ and $dz_{123}\in \Lambda^3\C^3$ correspond to the following octonions
 \be
 \Lambda^0\C^3 \ni \im 1 \leftrightarrow {\bf 1}+\im {\bf u}\in \Oc_\C, \qquad 
 \Lambda^3\C^3 \ni \im dz_{123} \leftrightarrow {\bf 1}-\im {\bf u}\in \Oc_\C.
 \ee
Taking this into account we get the following identification of particles with directions in $S_+$
 \be
\left( \begin{array}{c} \bar{\nu} \\ \bar{e} \end{array}\right) \leftrightarrow \left( \begin{array}{c} \bar{\nu} \\ \bar{e} \end{array}\right) \in S_+, \qquad \bar{\nu},\bar{e} \in {\rm Span}( {\bf 1}+\im{\bf u}), \\ 
  \left( \begin{array}{c} \bar{u} \\ \bar{d} \end{array}\right) \leftrightarrow \left( \begin{array}{c} \bar{u} \\ \bar{d} \end{array}\right)  \in S_+, \qquad \bar{u},\bar{d} \in {\rm Span}(  {\bf e}^1+\im {\bf e}^5, {\bf e}^2+\im {\bf e}^6, {\bf e}^3+\im {\bf e}^7),
 \ee
 and
  \be
 L = \left( \begin{array}{c} \nu \\ e \end{array}\right) \leftrightarrow  \left( \begin{array}{c} \nu \\ e \end{array}\right) \in S_+, \qquad \nu,e\in  {\rm Span}( {\bf 1}-\im{\bf u}),\\
 \nonumber
 Q = \left( \begin{array}{c} u \\ d \end{array}\right) \leftrightarrow  \left( \begin{array}{c} u \\ d \end{array}\right) \in S_+, \qquad u,d \in {\rm Span}(  {\bf e}^1-\im {\bf e}^5, {\bf e}^2-\im {\bf e}^6, {\bf e}^3-\im {\bf e}^7). 
 \ee 
  This gives the explicit dictionary identifying particles with vectors in $S_+$. 
  
  \section{Product structures and characterisation of $G_{\rm SM}$}
  \label{sec:new}
  
  To select a copy of the SM gauge group inside ${\rm Spin}(10)$ we needed to select a pure spinor, whose stabiliser gave us a copy of ${\rm SU}(5)$, as well as a J-compatible Pati-Salam subgroup, or alternatively a split $\C^5=\C^2\oplus\C^3$. This two pieces of data appear to be of different geometric origin, which is not satisfactory. In this Section we will use the fact that a J-compatible copy of Pati-Salam gauge group, or a split $\C^5=\C^2\oplus\C^3$ is the same as a $J$-compatible product structure. A $J$-compatible product structure can in turn be represented as the product of two commuting complex structures. This leads us to the main observation of this Section, which is that we can encode all the data required to choose a copy of the SM gauge group inside ${\rm Spin}(10)$ into two pure spinors of ${\rm Spin}(10)$ that are appropriately compatible.

  \subsection{A pair of commuting complex structures from pure spinors}
  \label{sec:pair-pure-spin}
  
  We have already seen in Theorem \ref{pure-10} how a pair of pure spinors of ${\rm Spin}(8)$ that are appropriately aligned gives two commuting complex structures on $\R^8=\Oc$. In the context of this theorem, the pure spinors are of opposite parity, and when the complex structures they define commute, make up a pure spinor of ${\rm Spin}(10)$. 
 
We now want to apply similar considerations in the context of ${\rm Spin}(10)$. Thus, we take a pair of pure spinors of ${\rm Spin}(10)$ $\Psi_{1,2}$, with their associated complex structures $J_1, J_2$. There is a choice to be made as to whether $\Psi_{1,2}$ are of the same parity or if their parities are opposite. Without a loss of generality we can assume that the parity is the same because if $\Psi_1\in S_+$ and $\Psi_2\in S_-$ then $R(\Psi_2)\in S_+$. So, we start by assuming that both $\Psi_{1,2}\in S_+$. 
  
As we already know from e.g. discussing surrounding (\ref{commuting-J}), the complex structures $J_{1,2}$ commute if and only if the eigenspace $E_1$ of eingevalue of $+\im$ of $J_1$ splits into the eigenspaces of $J_2$. Further, given a pair of pure spinors $\Psi_{1,2}\in S_+$, they share either $3$ or $1$ null direction. This happens in general, even in the case $J_{1,2}$ do not commute. The condition that $J_{1,2}$ commute is then that the complementary directions to those shared are also shared. It is best to consider the two possible situations that can arise in dimension ten separately.

Let us first consider the case when ${\rm dim}(M(\Psi_1)\cap M(\Psi_2))=1$. In order for the complex structures $J_{1,2}$ to commute the complementary null space $\overline{M(\Psi_2)}$ to e.g. $M(\Psi_2)$ must intersect with $M(\Psi_1)$ in precisely four null directions. This means that the following conditions must hold
\be\label{SU4}
H_{\Psi_1,\Psi_2} = {\rm SU}(4): \qquad B_0(\Psi_1, R(\Psi_2))=0, \qquad B_2(\Psi_1, R(\Psi_2))=0.
\ee
Because four null directions are shared in this case, it is clear that intersection of the ${\rm SU}(5)$ stabilisers of the spinors $\Psi_{1,2}$ is the subgroup ${\rm SU}(4)$, which is what we indicated in (\ref{SU4}). 

Let us now consider the other possible case when ${\rm dim}(M(\Psi_1)\cap M(\Psi_2))=3$. The complex structures $J_{1,2}$ will commute if in addition we have ${\rm dim}(M(\Psi_1)\cap \overline{M(\Psi_2)})=2$. This is the case when
\be\label{SU3-SU2}
H_{\Psi_1,\Psi_2} = {\rm SU}(3)\times{\rm SU}(2): \qquad B_1(\Psi_1, \Psi_2)=0, \qquad B_0(\Psi_1, R(\Psi_2))=0.
\ee
The intersection of the ${\rm SU}(5)$ stabilisers of $\Psi_{1,2}$, or equivalently their joint stabiliser, is in this case ${\rm SU}(3)\times{\rm SU}(2)$. As we know from the discussion in the preceding sections, it is the second one that is relevant for the symmetry breaking from ${\rm Spin}(10)$ to $G_{\rm SM}$, because the product structure $K=J_1 J_2$ is the one that gives the decomposition $\R^{10}=\R^6\oplus \R^4$. We can now rephrase the conditions (\ref{SU3-SU2}) as follows:
   \begin{proposition} All the conditions in the case (\ref{SU3-SU2}) can be summarised as the condition that an arbitrary linear combination of $\Psi_{1}$ and $\Psi_2$ is a pure spinor, and that $\Psi_{1,2}$ are orthogonal.
  \end{proposition}
  To see this we recall that $\Psi=a\Psi_1+b\Psi_2$ is a pure spinor when $B_1(\Psi,\Psi)=0$. Given that $\Psi_{1,2}$ are both pure, this condition is equivalent to $B_1(\Psi_1,\Psi_2)=0$. For the remaining conditions, $B_0(\Psi_1,R(\Psi_2))=0$ is just the statement that $\Psi_1\in S_+, R(\Psi_2)\in S_-$ have a vanishing pairing, and thus orthogonal. 
  
 We can use this result to provide a new characterisation of $G_{\rm SM}$ inside ${\rm Spin}(10)$:
 
\begin{manualtheorem}{A}
\label{thm:A} The SM gauge group $G_{\rm SM}\subset {\rm Spin}(10)$ is the group that stabilises a pure spinor $\Psi_1$ and projectively stabilises another pure spinor $\Psi_2$, with $\Psi_{1,2}$ orthogonal and such that an arbitrary linear combination of $\Psi_{1,2}$ is still pure. The direction $\Psi_1\in S_+$ is then identified with the sterile neutrino $\bar{\nu}$ direction, and the direction of $\Psi_2$ with the $\bar{e}$ direction. 
 \end{manualtheorem}

 \begin{remark} The condition that a spinor is stabilised decreases the rank of the gauge group by one. So, requiring one spinor to be stabilised reduces rank five to rank four, which is the rank of $G_{\rm SM}$. We cannot require another spinor to be stabilised if we want to reproduce $G_{\rm SM}$. At most, we can require it to be stabilised projectively, i.e. up to multiplication by a phase. This is precisely what is demanded by the above theorem.
 \end{remark}
  
  \subsection{The second pure spinor for $G_{\rm SM}$}

The above theorem follows from the discussion that preceded it. The only not obvious part of the statement is the fact that the spinor that is required to be stabilised projectively is to be identified with the $\bar{\nu}$ direction. Let us confirm this by a computation. 

Let us choose, as in the discussion in subsection \ref{sec:GG}, the first pure spinor $\Psi_1$ to be given by (\ref{psi-special}). Let us compute the second pure spinors $\Psi_2$ whose complex structure commutes with that of $\Psi_1$ and such that we get the split $\C^5=\C^3+\C^2$. We want the null vectors of this pure spinor to be
  \be
  e_1+\im e_5, \quad e_2+\im e_6, \quad e_3+\im e_7, \quad e_4-\im e_8, \quad e_9-\im e_{10},
  \ee
  so that the first 3 are shared with those in (\ref{eigendir}), and the last two are shared with the null directions of $R(\Psi_1)$. A moment of reflection shows that in terms of polyforms, while the first spinor $\Psi_1=1$ is a multiple of the identity polyform, the second spinor is a multiple of $dz_{45}$
  \be
  \Psi_1=1, \qquad \Psi_2=dz_{45}.
  \ee 
  Comparing with (\ref{sp-polyforms}) we see that this corresponds to 
  \be\label{two-spinors}
  \Psi_1 = \left( \begin{array}{c} {\bf 1}+\im {\bf u} \\ 0 \end{array}\right), \qquad  \Psi_2 = \left( \begin{array}{c} 0\\ {\bf 1}+\im {\bf u} \end{array}\right).
  \ee
  It is clear that $\Psi_2$ points in the direction of $\bar{e}$, as promised.
  
  So far we have considered the case when both $\Psi_{1,2}\in S_+$. It is also interesting to consider the possibility when the two spinors are of different parity. So, let us also consider 
  \be\label{psi2-bar}
  S_- \ni R(\Psi_2) = \left( \begin{array}{c} 0\\ {\bf 1}-\im {\bf u} \end{array}\right).
  \ee
It is interesting to translate this spinor into an odd polyform in $\Lambda^{odd}\C^5$, and thus to a direction in the particle space, as in (\ref{split-GG}). To this end, we recall from (\ref{psi-pm-10}) that a two-component column with octonionic entries that represents a negative parity spinor corresponds to the polyform $\gamma\wedge dz_5+ \delta$, where the complexified octonion $\gamma$ represents a polyform in $\Lambda^+(\C^4)$, and $\delta$ represents a polyform in $\Lambda^-(\C^4)$. In the case (\ref{psi2-bar}) we have $\delta= {\bf 1}-\im {\bf u}$. According to (\ref{polyforms-octonions-negative}) this corresponds to the polyform 
\be
R(\Psi_2) = \im dz_{123}.
\ee
Thus, according to (\ref{split-GG}), the polyform that corresponds to $R(\Psi_2)$ also represents the direction $\bar{e}$ in the particle space. This is of course the expected conclusion.

  \subsection{Explicit computation of the stabiliser}
  
 Even though Theorem \ref{thm:A} follows from considerations involving intersections of pure spinors, it is instructive to present an explicit verification. We consider the action of the Lie algebra $\mathfrak{spin}(10)$ and compute the stabiliser subalgebra of the two spinors (\ref{two-spinors}), keeping it in mind that the second pure spinor should only be stabilised projectively. The goal is to see the Lie algebra of $G_{\rm SM}$ arising. 
 
 We first consider the action of $\mathfrak{spin}(10)$, whose general element is given by (\ref{spin-10}) on $\Psi_1$. Requiring this action to vanish we get the equations  
 \be
 (A+\im a)({\bf 1}+\im{\bf u}) =0, \qquad (L_x+\im L_y)({\bf 1}+\im{\bf u}) =0.
 \ee 
 Writing down the the corresponding real and imaginary parts we get
 \be\label{eqs-psi-1}
 A{\bf 1}- a{\bf u} =0, \quad A{\bf u} +a{\bf 1}=0, \quad L_x{\bf 1} -L_y {\bf u}=0, \quad L_y {\bf 1} + L_x{\bf u}=0.
 \ee
 Let us start with the second pair of equations. They imply $x=y{\bf u}$, which thus links $x$ with $y$, thus imposing 8 conditions on the Lie algebra parameters. 
 
 The first equation, projected onto the ${\bf u}$ direction, imposes the condition
 \be\label{trace-eqn-psi-1}
 X^{15}+X^{26}+X^{37}-X^{84}+a=0.
 \ee
 Here $X^{IJ}$ are the parameters of the $\mathfrak{spin}(8)$ Lie algebra. The second equation, projected onto the direction ${\bf 1}$, imposes the same equation. There are six more equations in both the first and the second equation. Their sums and differences are then the conditions of the type $X^{81}+X^{45}=0, X^{27}-X^{36}=0$. Altogether, the equations in (\ref{eqs-psi-1}) impose $8+1+12=21$ condition on $45$ parameters, which gives 24 remaining parameters, which is the expected dimension of $\mathfrak{su}(5)$. 
 
 We now impose the condition that $\Psi_2$ is projectively stabilised. This gives the equations 
 \be
 (-L_{\bar{x}}+\im L_{\bar{y}})({\bf 1}+\im{\bf u})=0, \qquad (A'-\im a)({\bf 1}+\im{\bf u}) = \im b ({\bf 1}+\im{\bf u}),
 \ee 
 where $b$ is an arbitrary parameter. The real and imaginary parts of these equations are
 \be
 L_{\bar{x}} {\bf 1} + L_{\bar{y}}{\bf u}=0, \quad  -L_{\bar{x}} {\bf u} + L_{\bar{y}}{\bf 1}=0, \quad
 A'{\bf 1} + a{\bf u} = -b {\bf u}, \quad A'{\bf u} -a{\bf 1} = b{\bf 1}.
 \ee
 The first pair implies $x={\bf u}y$. The third equation projected onto the ${\bf u}$ direction gives
 \be\label{trace-eqn-psi-2}
- X^{15}-X^{26}-X^{37}-X^{84}+a=-b.
\ee
The fourth equation projected into ${\bf 1}$ direction gives the same condition. The rest of the equations in the last pair imply 12 equations of the sort $X^{81}-X^{45}=0, X^{27}-X^{36}=0$. All in all, given the fact that $b$ is arbitrary and (\ref{trace-eqn-psi-2}) just determines $b$, the conditions that $\Psi_2$ is projectively stabilised reduces the group to a copy of ${\rm U}(5)$. 

Let us now consider the intersection of the ${\rm SU}(5)$ that stabilises $\Psi_1$ and ${\rm U}(5)$ that projectively stabilises $\Psi_2$. The equations $x=y{\bf u}, x={\bf u} y$ imply that $y$ commutes with ${\bf u}$. This means that
\be\label{xy}
y=p {\bf 1}+q {\bf u} \quad p,q\in \R, \quad \text{and}\quad  x=y{\bf u}={\bf u}y.
\ee
The equations $X^{81}+X^{45}=0, X^{81}-X^{45}=0$ imply $X^{81}=X^{45}=0$. There are $6+6$ equations like these, that reduce $\mathfrak{spin}(8)$ to $\mathfrak{spin}(6)$ acting in the directions $1,2,3,5,6,7$. Moreover, there remain 6 more equations of the type $X^{27}-X^{36}=0$, which further reduce $\mathfrak{spin}(6)=\mathfrak{su}(4)$ to $\mathfrak{u}(3)$. 
  
Let us now consider the remaining transformations  in the $4,8,9,10$ directions. The remaining parameters are $X^{84}, a$ as well as two parameters $p,q$ in $x,y$. The parameters $X^{84},a$ can be parametrised by their sum and difference. The difference is constrained by (\ref{trace-eqn-psi-1}). To sum, together with $\xi,\zeta$ combine into a copy of ${\rm SU}(2)$. Indeed, let us consider the Lie algebra element with $X^{84}=a$ and $x,y$ as in (\ref{xy}). A simple computation shows that this Lie algebra element can be written as
\be
\left(\begin{array}{cc} a & q+\im p \\ q-\im p & - a \end{array}\right) (\im + L_{\bf u}).
\ee
 This Hermitian tracefree $2\times 2$ matrix gives a copy of Lie algebra $\mathfrak{su}(2)$ acting non-trivially on the $-\im$ eigenspace of the complex structure $J_S=-L_{\bf u}$. Thus, this is the Lie algebra of ${\rm SU}(2)$ acting non-trivially on the second term in (\ref{split-Sp}), and so is the weak ${\rm SU}(2)$.

The difference $X^{84}-a$ gives an additional copy of ${\rm U}(1)$ that extends this $\mathfrak{su}(2)$ to $\mathfrak{u}(2)$, and then the condition (\ref{trace-eqn-psi-1})  relates the $\mathfrak{u}(1)$ generators of $\mathfrak{u}(3)$ and $\mathfrak{u}(2)$, producing the Lie algebra of the SM gauge group. 
  
 It can be checked that with our normalisations the $Y$ generator of ${\rm U}(1)_Y$ is given by
 \be
 Y = \frac{1}{6}( \Gamma^1 \Gamma^5+ \Gamma^2 \Gamma^6+ \Gamma^3 \Gamma^7) + \frac{1}{2} 
 (\Gamma^8 \Gamma^4- \Gamma^9 \Gamma^{10}).
 \ee
 
\section{Discussion}
  
The natural question to ask is what our main new result means in terms of possible model building. This is a geometry paper rather than one on phenomenology, and so it would not be appropriate to attempt to study any concrete models of the symmetry breaking. The most we can do is describe a possible Lagrangian that can realise the geometric ideas of this paper. 

It is clear that the Higgs Lagrangian must involve two spinor Higgs fields. Let us denote them by $\Phi_1, \Phi_2$. We can then write the usual kinetic term for $\Phi_1$ that is built using the ${\rm Spin}(10)$ covariant derivative
\be\label{kin-term}
{\cal L}_1 = \frac{1}{2}\langle R((\partial_\mu + {\bf A}_\mu) \Phi_1), (\partial_\mu + {\bf A}_\mu) \Phi_1\rangle.
\ee
Here $\bf A$ is the ${\rm Spin}(10)$ gauge field, valued in the same spinor representation $S_+$ where $\Phi_1$ lives, and $R$ is the anti-linear operator that maps $S_+\to S_-$ and so allows the spinor inner product to be taken. 

The (quartic) potential for $\Phi_1$ can easily be designed to be such that it is minimised by a configuration of $\Phi_1$ in which it is a pure spinor. Indeed, this is clear e.g. from the expression (\ref{X2-Y2}) for the norm squared of a vector that must vanish when the spinor is pure. This makes it clear that any positive multiple of $\langle R(\Phi_1),\Phi_1\rangle^2 - 4I(\Phi^1)$ is a non-negative potential that attains zero values if and only if the spinor $\Phi_1$ is pure. We can write this potential explicitly as 
\be
V(\Phi_1) = \frac{\kappa_1}{2} \langle \Phi_1, \Gamma_I \Phi_1\rangle \overline{\langle \Phi_1, \Gamma^I \Phi_1\rangle}.
\ee
Here $\Gamma_I, I=1,\ldots,10$ are the ${\rm Cl}(10)$ gamma-matrices, and the overline denotes the complex conjugation. In spite of this being a homogeneous degree four potential, it is minimised (with zero value at the minimum) when $\Phi_1$ is a pure spinor. The spinor $\Phi_1$ minimising this potential breaks ${\rm Spin}(10)$ to ${\rm SU}(5)$, and gives masses through (\ref{kin-term}) to all the gauge fields that are not in the ${\rm SU}(5)$ that stabilises $\Phi_1$. Note, however, that this potential does not determine the value of the VEV of $\Phi_1$, it just requires it to point in the direction of a pure spinor. In case one wishes to set the VEV of $\Phi_1$ with a potential, this can easily be achieved by adding the usual Higgs-type potential
\be
V'(\Phi_1) = \frac{\kappa_1'}{2}( \langle R(\Phi_1),\Phi_1\rangle - v_1)^2.
\ee

The situation with the second Higgs field $\Phi_2$ is more subtle. Thus, as the Theorem \ref{thm:A} states, we should only require it to be stabilised modulo phase. To satisfy this we introduce an additional auxiliary ${\rm U}(1)$ gauge field $\omega_\mu$ that only couples to $\Phi_2$. The kinetic term is then
\be
{\cal L}_2 = \frac{1}{2}\langle R((\partial_\mu + {\bf A}_\mu+\omega_\mu) \Phi_2), (\partial_\mu + {\bf A}_\mu+\omega_\mu) \Phi_2\rangle.
\ee
We supplement this with the potential term for $\Phi_2$ that requires $\Phi_2$ to be a pure spinor, similar to $\Phi_1$
\be
V(\Phi_2) = \frac{\kappa_2}{2} \langle \Phi_2, \Gamma_I \Phi_2\rangle \overline{\langle \Phi_2, \Gamma^I \Phi_2\rangle},
\ee
and a potential that sets the VEV of this Higgs field
\be
V'(\Phi_2) = \frac{\kappa_2'}{2}( \langle R(\Phi_2),\Phi_2\rangle - v_2)^2.
\ee

The last bit of the Lagrangian to be discussed is the potential that depends on both $\Phi_1,\Phi_2$. It is clear one part of this potential is one that makes an arbitrary linear combination of $\Phi_1,\Phi_2$ pure when both are pure spinors. When $\Phi_1,\Phi_2$ are both pure, the complex $\R^{10}$ vector that is constructed from their linear combination is a multiple of $\langle \Phi_1,\Gamma_I \Phi_2\rangle$. And so the norm squared of this complex vector is zero if and only if a linear combination of the pure spinors $\Phi_1,\Phi_2$ is also pure. The required potential is then
\be
V(\Phi_1,\Phi_2) = \frac{\kappa_{12}}{2} \langle \Phi_1, \Gamma_I \Phi_2\rangle \overline{\langle \Phi_1, \Gamma^I \Phi_2\rangle}.
\ee
However, we should also require that $\Phi_1, R(\Phi_2)$ are orthogonal. The potential that achieves this is
\be
V'(\Phi_1,\Phi_2) = \frac{\kappa'_{12}}{2} \langle \Phi_1, R(\Phi_2)\rangle \overline{\langle \Phi_1, R( \Phi_2)\rangle}.
\ee
It is clear that the configuration of   $\Phi_1,\Phi_2$ that minimises all the parts of the potential is as required by the Theorem \ref{thm:A}, and that the gauge fields that remain massless are those in a copy of $G_{\rm SM}$ inside ${\rm Spin}(10)$. This gives a concrete theoretical model that realises the geometric ideas described in the present paper. A phenomenological study of this model will be attempted elsewhere.

  \section*{Acknowledgements}
The author is grateful to Ivan Todorov and Niren Bhoja for collaboration on the initial stages of this project. I am also grateful to F. Reese Harvey, Nigel Hitchin and Michal Malinsky for correspondence, and to Pietro Antonio Grassi for a discussion about pure spinors in 10D and references. 


\end{document}